\documentclass[prd,showpacs]{revtex4}
\usepackage{bm}
\usepackage{graphicx}
\usepackage{amsmath,amsfonts,amsbsy}

\def \pd {\partial}
\def \d3k {\frac{d^3k}{(2\pi)^3}}
\def \v#1{{\bm #1}}
\def \in {{\rm in}}
\def \be {\begin{equation}}
\def \ee {\end{equation}}
\newcommand{\bal}{\begin{align}}
\newcommand{\eal}{\end{align}}

\begin{document}

\title{Motion of Classical Impurities in the Homogeneous Bose-Einstein Condensate}
\author{Jun Suzuki}
\thanks{Present address: Department of Physics, National University of Singapore, Singapore 117542, Singapore}
\email{physj@nus.edu.sg}
\affiliation{Department of Physics and Astronomy, University of South Carolina, Columbia, SC 29208}
\date{June 6, 2005}

\begin{abstract}
Motion of classical point-like impurities in the homogeneous
Einstein condensate of bosons is studied in the framework of
second quantization method. A toy model is proposed and its
general solution within the Bogoliubov approximation is obtained.
The effective Minkowski space-time structure arises naturally in
this non-relativistic quantum many-body system in the low energy
regime. This is shown to be true in this model. Several examples
are discussed in order to illustrate our model. The homogeneous
condensate produces an effective Yukawa type attractive force
between impurities sitting in condensate. Landau's criterion is
naturally derived in a case of linear motion of impurity. The
analytic expressions for spectra of Bogoliubov excitations
produced by the accelerated motions of impurities are obtained. A
quick look at the analytic expression reveals that the spectrum of
gapless excitations emitted by the linearly accelerated impurity
{\it is not thermal}. If the homogeneous condensate is the 
physically correct model for Minkowski space-time then it follows 
that the apparent thermal response of the simple linearly 
accelerated detector models may be the result of improper regularization.
\end{abstract}

\maketitle

\section{Introduction}
In 1924 Einstein predicted the very peculiar behavior of an ideal
gas of non-relativistic particles obeying a new statistics
\cite{einstein}. This new statistics (Bose-Einstein statistics)
was proposed by Bose to explain the Planck distribution for the
black body radiation. Einstein found that most of the population
of the ideal gas would occupy the ground state below the critical
temperature. This phenomenon is somewhat counterintuitive, and has
no place in classical statistics, but it revealed the quantum
nature of many particle systems. The phenomenon is now known as
Bose-Einstein condensation, referring to the Einstein condensation
(entartung) in systems having the Bose-Einstein statistics. In
1995 Bose-Einstein condensate (BEC) was finally created in
laboratories in several alkali vapors \cite{bec1,bec2,bec3}. 
The essential difference besides developments in cooling techniques is
an additional magnetic trapping to confine the cold atoms within a
small region.  This celebrated experimental success also revived
theoretical interests in BECs with additional trapping potentials
\cite{bec}.

A decade after the experimental realization of BEC,
this field is still very active, both experimentally and theoretically.
Recently much interest has been growing in cosmology and gravitational physics.
In these fields it is very difficult to do experiments in laboratories, and
experimental tests are usually based on astrophysical observations.
There are several proposals to resolve this discrepancy,
which discuss ``analogies" between real gravity and BECs \cite{analog}.
In their proposals the homogeneous BEC serves as a Minkowski space-time and
the deviations from the flat metric tensor arise as a so called ``acoustic metric" in
inhomogeneous BECs. In this way BECs may mimic ``analog" gravities.
Hence ``effective" gravities in laboratories are attainable to test the theories.

On the other hand, there is another direction of research where
gravitation is indeed considered as an emergent phenomenon arising
from collective excitations in quantum many-body systems
\cite{mazur0, mazur1, mazur2, mazur3, mazur4, mazur5, mazur6, laughlin,
chapline}. In these proposals a physical vacuum is considered as a
real condensed state which contains a huge number of heavy bosons.
The Minkowski vacuum is assumed to be the homogeneous BEC. The low
energy vacuum structure of such an medium is then described by the
relativistic quantum field theory. In this sense the relativistic
quantum field theory is an effective theory which works only in
the low energy scale. It then follows that the relativistic
quantum field theory breaks down at certain energy scale. Many
difficulties of the relativistic quantum field theory seem to
occur as a result of an extrapolation and the usual regularization
techniques corresponding to unphysical degrees of freedom.

The idea of the space-time medium for the propagation of light in quantum mechanics was
introduced by Dirac \cite{dirac1, dirac2}. The old idea of an \ae ther was abandoned as a result of success of
special relativity and by the Michelson-Morley experiment before quantum mechanics.
Recent precise experiment shows that the upper limit for the anisotropy of the velocity of light is $10^{-15}$ \cite{michelson}.
Therefore it is an experimental fact that the velocity of light is isotropic.
Dirac revisited the possibility of keeping the idea of an \ae ther in quantum mechanics.
He showed that the isotropy of the velocity of light
does not conflict with the idea of the space-time medium
which is responsible for the propagation of light.
From this point of view, a quantum many-body system could be a candidate
since the velocity of sound is isotropic in several BECs and in $^4$He.

It is of fundamental interest to study the motion of impurities in BECs.
By impurities we mean any bosons or fermions which can be distinguishable from condensed bosons,
and the total number of impurities $N_{imp}$ is much less than that of condensed bosons $N$ ($N\gg N_{imp}$).
The general solution to this problem has not been found so far, and we are unlikely
to obtain it within our current understanding in many-body problems.
Several partial results are obtained recently from a condensed matter
physics point of view \cite{meystre,kovrizhin,ms,astracharchik,mazets,timmermans,jun}.
This problem is not only interesting in a pure academic sense.
It is also very important for practical reasons to take into account the effects of impurities in experiments.

The purpose of this paper is to provide a small step toward solution of this problem.
We propose a toy model describing the motion of classical impurities in
the homogeneous BEC in the framework of the second quantization method.
To this end we employ the Bogoliubov model for weakly interacting massive bosons
in a dilute regime \cite{bogoliubov, bogoliubov2, bru}.
We then add an interaction Hamiltonian which takes into account the local contact interaction between bosons and impurities.
Specifically, we focus on the motion of classical point-like impurities moving along given trajectories.
We do not discuss back reaction on the impurities, and their dynamics mainly.
Our main interest is to find out how the homogeneous condensate
responds to the motion of impurities, and how it tries to maintain its condensed state in the presence of impurities.
We also want to examine the hypothesis postulated in previous studies,
stating that the homogeneous BEC is the Minkowski vacuum.
Especially important is finding out how the effective Minkowski space-time structure emerges from
the non-relativistic many-body system. Is the motion of impurities in the homogeneous BEC
exactly corresponding to the motion of test particles in the Minkowski space-time?
Suppose that the homogeneous condensate happens to be unstable against the motion of impurities
and can be easily destroyed by them. If it happens, then their hypothesis should be rejected.
This is because the Minkowski vacuum has to be very stable, as is well known.
We will show that the homogeneous condensate is very robust in a large $N$ limit
and can give rise to macroscopic effects which could be observed.

After we demonstrate that the homogeneous condensate has similar
properties to the Minkowski vacuum, we will study several examples
to illustrate our model. Particularly, the accelerated motion of
impurities in BEC is of interest. This situation is analogous to
the case of accelerated moving detectors in the Minkowski
space-time. According to the standard description of quantum field
theory in the curved space-time, a uniformly accelerated detector
in the Minkowski space-time would detect a thermal spectrum. In
other words, observers ``feel hot" when they are accelerated in
the Minkowski space-time. The temperature $T_a$ which such an observer
would measure is given by the formula, 
\be 
T_a=\frac{\hbar a}{2\pi k_B c}. 
\ee 
Here $a$ is the acceleration constant of the observer at
an instantaneous rest frame, and $k_B$ is the Boltzmann constant.
This thermal effect is generally referred to as the ``Unruh effect"
\cite{unruh}. Thus, it has long been believed that an empty
Minkowski vacuum would serve as a thermal bath for accelerated
observers. In spite of  the absence of experimental confirmation,
this mathematical consequence has been accepted for over three
decades.  There is sometimes an impression that this thermal
character is equivalent to quantum field theory itself
\cite{unruh2}. Recently, Belinskii and his collaborators revisited
the Unruh effect. They claim that the thermal character of the
Minkowski space is an artifact of an invalid mathematical
procedure repeatedly used in the discussion of the Unruh effect
\cite{belinskii,belinskii2, belinskii3}. The problem remains open.

Therefore it is very interesting to examine the accelerated motion of impurities in the homogeneous condensate.
In the model presented in this paper, the thermal spectrum is not found for cases such as a uniform accelerated motion,
nor a uniform circular motion.  
This paper is organized as follows.  
In Sec.~II we propose a simple model Hamiltonian describing a point-like classical impurity in BEC, 
and we obtain the general solution within the Bogoliubov approximation. 
In order to keep this paper self-contained, 
we give the review on the number conserving Bogoliubov model briefly in this section. 
We then derive formulae for the occupation number and the energy dissipation in Sec.~III.  
The general solution obtained in Sec.~II provides time dependent behavior of various physical quantities 
such as the density, the current density, the stress energy momentum complex. 
Deviations of these quantities from the homogeneous condensate are also discussed. 
Low energy behavior of these quantities suggest that the Minkowski space-time 
be an emergent space-time arising from non-relativistic quantum many-body system, i.e. the homogeneous BEC. 
In Sec.~IV we study several examples to illustrate our model and its general solution. 
We first provide the general argument for the derivation of Landau's criterion based on 
the method of stationary phase. Next we obtain an attractive Yukawa type force between 
two stationary impurities sitting in the BEC. Then, three particular examples for the motion of 
impurity are studied; a linear motion with a constant velocity, a uniformly accelerated motion, 
and a circular motion with a constant angular velocity. 
We obtained analytic expressions for these cases with thorough analyses of them. 
Sec.~V provides brief discussion for the dynamics of a quantum impurity in our model. 
We obtained a closed equation for the motion of single, point-like quantum impurity. 
In Sec.~VI we also give an alternative model for the motion of impurities based on the Galilean invariance. 
The current-current type interaction model is proposed and solved within the same approximation done 
in Sec.~II.  We summarize our results and give a discussion on our results in Sec.~VII. 
Appendix contains a supplement for the matin text.

\section{Model Hamiltonian and its Diagonalization}
\subsection{The Model}
The Bogoliubov model Hamiltonian for weakly interacting bosons of mass $M$ is
\be \label{h0}
\hat{H}_0 (t) = \int\!d^3x\;\hat{\psi}^{\dagger} (\v{ x},t) (-\frac{\hbar ^2 \v{ \nabla} ^2}{2 M})\hat{\psi} (\v{ x},t) \\
+ \frac 12  \int\!d^3x\!\!\int\!d^3x'\hat{\psi}^{\dagger} (\v{ x},t) \hat{\psi}^{\dagger}
(\v{ x'},t)V(\v x-\v x') \hat{\psi} (\v{ x},t) \hat{\psi} (\v{ x'},t) .
\ee
The field operators $ \hat{\psi}^{\dagger} (\v{ x},t)$ and $ \hat{\psi} (\v{ x},t)$ in the Heisenberg picture
satisfy the equal time canonical commutation relations :
\be
\begin{split}
[ \hat{\psi} (\v{ x},t),  \hat{\psi} ^{\dagger}(\v{ x'},t)] &= \delta (\v x -\v x'),\\
[\hat{\psi} (\v{ x},t),  \hat{\psi} (\v{ x'},t)]&=[\hat{\psi}  ^{\dagger}(\v{ x},t),  \hat{\psi} ^{\dagger} (\v{ x'},t)]=0.
\end{split}
\ee
The general two-body interaction term $V(\v x-\v x') $ can be approximated by a contact interaction
$g \delta(\v x-\v x')$ in the low energy region. This approximation will be used throughout the paper
in order to get analytic results, but extension to the general two-body interaction is quite straightforward \cite{comment31}.
The coupling constant $g$ for the approximated interaction between bosons is
expressed in terms of the s-wave scattering length $a_s$; $g=4 \pi a_s \hbar ^2/M$.
We assume the repulsive interaction $g>0$. The condition of diluteness $na_s^3\ll1$ is also assumed. Here
$n$ is the number density of massive bosons.
We consider the following interaction Hamiltonian between bosons and the distribution of classical impurities $\rho_c (\v{ x},t)$
taking into account the local interactions between them.
\be \label{hi}
\hat{H}_{{\rm I}} (t) = \lambda  \int d^3x\;\rho _c (\v{ x},t) \hat{\psi}^{\dagger} (\v{ x},t) \hat{\psi} (\v{ x},t) .
\ee

For a point-like classical impurity moving along a given trajectory $\v{ \zeta}(t)$, $\rho_c (\v{ x},t)$ is
written using the Dirac delta distribution as $\rho_c (\v{ x},t)=\delta (\v x -\v{ \zeta}(t))$.
The coupling constant $\lambda$ is expressed in terms of the s-wave scattering length $b_s$;
$\lambda =2 \pi b_s^i \hbar ^2/\tilde{m}$.
Here $\tilde{m}$ is the reduced mass of the boson and the impurity.
Extension to an arbitrary number of point-like impurities case is obtained by a replacement
$\lambda \rho_c(\v{ x},t)\to \sum_i \lambda _i \delta (\v x -\v{ \zeta}_i(t))$.

The momentum operator of the bosons
\be
 \hat{\v P}=\int d^3x\;\hat{\psi}^{\dagger} (\v{ x},t)  (-i \hbar \v \nabla) \hat{\psi} (\v{ x},t),
 \ee
commutes with the Hamiltonian (\ref{h0}). However, the momentum operator
does not commute with the interaction Hamiltonian (\ref{hi}).
This is contrary to the fundamental requirement of translational invariance.
This should be understood as follows. Let us consider a quantum mechanical impurity
described by its position and momentum operator $\hat{\v q}(t)$ and $\hat{\v p}(t)$ respectively.
The interaction Hamiltonian in this case reads
\be
\hat{H}_{{\rm I}} (t) = \lambda  \int d^3x\;\delta(\v x-\hat{\v q}(t))\hat{\psi}^{\dagger} (\v{ x},t) \hat{\psi} (\v{ x},t) .
\ee
Then the total momentum operator $ \hat{\v P}+ \hat{\v p}$ commutes with this interaction Hamiltonian.
The non-conserving momentum is now understood as an approximation of replacing q-number with c-number impurity
trajectory $\v \zeta(t)$. This approximation is valid as long as only slowly moving classical impurities are concerned.
The quantum impurity will be discussed later 
in Sec.~VI, where the closed equation is obtained for the quantum impurity. 

We expand the field operators in terms of the plane wave basis with
periodic boundary conditions in a finite size box $V=L^3$ as
\be \label{h1}
\hat{H}_0 (t)= \sum _{\v k} \epsilon _{k} \hat{a}_{\v k}^{\dagger}(t)\hat{a}_{\v k}(t)
+ \frac{g}{2V} \sum _{\v k,\v k',\v q} \hat{a}_{\v k+\v q}^{\dagger}(t)\hat{a}_{\v k'-\v q}^{\dagger}(t)\hat{a}_{\v k}(t)\hat{a}_{\v k'}(t)
\ee
and
\be \label{hi2}
\hat{H}_{{\rm I}}(t) = \frac{\lambda}{V} \sum _{\v k,\v k'} \tilde {\rho} _{\v k'}(t) \hat{a}_{\v k+\v k '}^{\dagger}(t)\hat{a}_{\v k}(t) ,
\ee
where $ \epsilon _{k}=\hbar ^2 \v k^2/2M$ is the free kinetic energy of bosons.
Here the summation is taken over integers $n_i$ for $\v k=2\pi(n_x,n_y,n_z)/L$.
In the above expression, we introduced the Fourier transform of the impurity distribution :
\be
\tilde{\rho} _{\v k}(t) =\int d^3 x \;\rho_c(\v x, t)e^{-i\v k \cdot \v x}.
\ee
By definition $\tilde{\rho} _{\v k=0}(t) =\int d^3 x \rho_c(\v x, t)=1$.

The creation and the annihilation operators $\hat{a} ^{\dagger}_{\v k}(t)$, $\hat{a}_{\v k}(t)$
satisfy the equal time canonical commutation relations :
\be
\begin{split}
[\hat{a}_{\v k}(t), \hat{a} ^{\dagger}_{\v k'}(t)] & = \delta_{\v k,\v k'}
,\\ [\hat{a}_{\v k}(t), \hat{a} _{\v k'}(t)] &=[ \hat{a} ^{\dagger}_{\v k}(t), \hat{a} ^{\dagger}_{\v k'}(t)]=0 ,
\end{split}
\ee
where $ \delta_{\v k,\v k'} =\delta_{n_x,n'_{x}}\delta_{n_y,n'_{y}}\delta_{n_z,n'_{z}}$
with $\delta_{n,n'}$ the Kronecker delta.
In the following discussion, time dependence of these operators for simplicity will be suppressed.

Later we will take the thermodynamic limit in which the volume $V$ and the number of bosons $N$ become
infinite, keeping the density $n = N/V$ fixed. In this limit the summation is replaced with
the integral
\be
\lim_{V\to \infty} \frac1V \sum_{\v k}= (2\pi)^{-3} \int d^3 k.
\ee
The creation and annihilation operators are relabeled by the continuous wave numbers as
\be
\lim_{V\to \infty} \sqrt{V}\hat{a} ^{\dagger}_{\v k}=(2\pi)^{3/2}\hat{a} ^{\dagger}(\v k),\;
\lim_{V\to \infty} \sqrt{V}\hat{a} _{\v k}=(2\pi)^{3/2}\hat{a}(\v k).
\ee
Then, $\hat{a} ^{\dagger}(\v k)$ and $\hat{a}(\v k)$ satisfy the canonical commutation relations:
\be
\begin{split}
[\hat{a}(\v k),\;\hat{a} ^{\dagger}(\v k')]  &= \delta(\v k-\v k'), \\
\ [\hat{a}(\v k),\;\hat{a} (\v k')] &=[ \hat{a} ^{\dagger}(\v k),\;\hat{a} ^{\dagger}(\v k')]=0.
\end{split}
\ee
In the following we are interested in
the effects of impurities to the homogeneous condensate in the large $N$ limit.  We will investigate the first order
correction on the homogeneous condensate.

We follow Bogoliubov's treatment \cite{bogoliubov, bogoliubov2, bru} to simplify
the full Hamiltonian $\hat{H}_0 +\hat{H}_{{\rm I}} $.
First we separate the second term in (\ref{h1}) into two parts, one is the forward scattering term with
zero momentum transferred $\hat{U}_0$ and the other with non zero momentum transferred $\hat{U}'$.
$\hat{U}_0$ and $\hat{U}'$ are
\be
\hat{U}_0  = \frac{g}{2V} \sum _{\v k,\v k'} \hat{a}_{\v k}^{\dagger}\hat{a}_{\v k'}^{\dagger}\hat{a}_{\v k}\hat{a}_{\v k'} 
= \frac{g}{2V} (\hat{N}^2-\hat{N}),
\ee
where $\hat{N} = \sum _{\v k}  \hat{a}_{\v k}^{\dagger} \hat{a}_{\v k}$ is the total number operator, and
\be
\hat{U}'= \frac{g}{2V} \sum _{\v k,\v k',\v q \neq 0} \hat{a}_{\v k+\v q}^{\dagger}
\hat{a}_{\v k'-\v q}^{\dagger}   \hat{a}_{\v k}   \hat{a}_{\v k'} .
\ee
The second term $\hat{U}'$ is split into three parts depending on the number of the
zero mode creation and annihilation operators as $\hat{U}'=\hat{U}^{(2)} +\hat{U}^{(1)} +\hat{U}^{(0)}$ :
\begin{align}
\hat{U}^{(2)} & = \frac gV \hat{a} ^{\dagger}_{0}\hat{a}_{0}\sideset{}{'}\sum \hat{a}_{\v k}^{\dagger} \hat{a}_{\v k}
+\frac{g}{2V} (\hat{a}_{0}\hat{a}_{0}\sideset{}{'}\sum \hat{a}_{\v k}^{\dagger} \hat{a}^{\dagger}_{- \v k} + {\rm h.c.}),\\
\hat{U}^{(1)} & = \frac gV (\hat{a}_{0}\sideset{}{'}\sum \hat{a}_{\v k+ \v k'}^{\dagger} \hat{a}_{\v k} \hat{a}_{\v k'} + {\rm h.c.} ),\\
\hat{U}^{(0)} & = \frac{g}{2V} \sideset{}{'}\sum \hat{a}_{\v k+\v q}^{\dagger}   \hat{a}_{\v k'-\v q}^{\dagger}   \hat{a}_{\v k}   \hat{a}_{\v k'} ,
\end{align}
where the prime over the summation is used to indicate the omission of the zero mode for the summation.
In the large $N$ approximation, we keep the $\hat{U}^{(2)}$ term only.  The rest contributes only in higher order corrections.

Next we eliminate the zero mode creation and annihilation
operators as follows \cite{comment2}. Define a self-adjoint operator $\hat{\beta}
_0$ by \be \label{beta} \hat{\beta}_0 = (\hat{N}_0 +1)^{-1/2}
\hat{a}_0 ,\ \hat{N}_0 = \hat{a} ^{\dagger}_{0}\hat{a}_{0} . \ee
This operator $\hat{\beta}_0 $ and its hermite conjugate satisfy
$\hat{\beta} _{0}\hat{\beta}^{\dagger}_{0}=1$ and $\hat{\beta}
^{\dagger}_{0}\hat{\beta}_{0}=1-\hat{\Pi}_0$, where
$\hat{\Pi}_0=|N_0=0\rangle\langle N_0=0|$ is the projection
operator onto the state $N_0 =0$. In the presence of the
homogeneous BEC for the zero mode, we can safely exclude the state
$N_0 =0$. We then approximate $\hat{\beta}
^{\dagger}_{0}\hat{\beta}_{0} \simeq 1$, i.e. $\hat{\beta}_0$ is
an almost unitary operator, and $[ \hat{\beta} _{0},
\hat{\beta}^{\dagger}_{0}]\simeq 0 $ holds. More precisely this
approximation holds between matrix elements when we compute
expectation values. Then $\hat{a}^{\dagger}_0=\sqrt{\hat{N}_0}
\hat{\beta}^{\dagger}_0$ and $\hat{a}_0= \hat{\beta}_0
\sqrt{\hat{N}_0} $ allow us to eliminate the bare zero mode
operators $\hat{a}^{\dagger}_0$ and $\hat{a}_0$. We introduce a
new set of the creation and the annihilation operators
$\hat{\alpha}^{\dagger}_{\v k}$ and $\hat{\alpha}_{\v k}$ for
non-zero momentum by \be \hat{\alpha}^{\dagger}_{\v k} = \hat{a}
^{\dagger} _{\v k}  \hat{\beta}_0 , \ \hat{\alpha}_{\v k} =
\hat{\beta}^{\dagger}_0 \hat{a} _{\v k} \quad {\rm for}\  \v k
\neq 0, \ee which satisfy the equal time canonical commutation
relations :
\be
\begin{split}
[\hat{\alpha}_{\v k}(t), \hat{\alpha} ^{\dagger}_{\v k'}(t)] & = \delta_{\v k,\v k'}
- \hat{\alpha}_{\v k'} \hat{\alpha} ^{\dagger}_{\v k} \hat{\Pi}_0 \simeq \delta_{\v k,\v k'}  ,\\
 [\hat{\alpha}_{\v k}(t), \hat{\alpha} _{\v k'}(t)] &
 =[ \hat{\alpha} ^{\dagger}_{\v k}(t), \hat{\alpha} ^{\dagger}_{\v k'}(t)]=0 .
\end{split}
\ee
The vacuum state of the Fock space ${\cal H}_0$ for the new set of the creation and the annihilation operators
is the same as before, and it is defined by
\be
\hat{\alpha}_{\v k} |0 \rangle =0 \quad {\rm for\ all}\ \v k \neq 0.
\ee
It should be noted that these composite creation and annihilation operators have the following properties :
\be
\hat{\alpha} ^{\dagger}_{\v k} |N_0, N_{\v 1}, \cdots, N_{\v k}, \cdots \rangle
= \sqrt{N_{\v k+1}}  |N_0-1, N_{\v 1}, \cdots, N_{\v k}+1, \cdots \rangle ,
\ee
and
\be
\hat{\alpha}_{\v k} |N_0, N_{\v 1}, \cdots, N_{\v k}, \cdots \rangle
=\sqrt{N_{\v k}}  |N_0+1, N_{\v 1}, \cdots, N_{\v k}-1, \cdots \rangle ,
\ee
where $N_{\v k}$ is the number of particles occupying the mode $\v k $ in the plane wave basis.
In other words, the total number of particles is conserved after we apply these operators.

We express the Hamiltonian (\ref{h1}) in terms of these new operators, $\hat{N}_0$, and $\hat{N}$,
\be \label{h2}
\hat{H}_0 =\frac{g}{2V} (\hat{N}^2-\hat{N})
+\sideset{}{'}\sum (\epsilon _{k}+\frac gV \hat{N}_0) \hat{\alpha}_{\v k}^{\dagger}\hat{\alpha}_{\v k}
+ \frac{g}{2V} ( \sqrt{\hat{N}_0+2}\sqrt{\hat{N}_0+1} \sideset{}{'}\sum
\hat{\alpha}_{\v k}^{\dagger}\hat{\alpha}^{\dagger}_{-\v k} + {\rm h.c.}).
\ee
Similarly for the interaction Hamiltonian (\ref{hi2}),
\be \label{hi3}
\hat{H}_{{\rm I}} = \frac {\lambda}{V} \hat{N}_0
+ \frac {\lambda}{V} (\sqrt{\hat{N}_0+1} \sideset{}{'}\sum \tilde{\rho} _{\v k}(t) \hat{\alpha}_{\v k}^{\dagger} +{\rm h.c.})
+ \frac {\lambda}{V} \sideset{}{'} \sum \tilde{\rho} _{\v k'}(t) \hat{\alpha}_{\v k+\v k '}^{\dagger}  \hat{\alpha}_{\v k}.
\ee

We note that the total number operator
$\hat{N}=\hat{N}_0+\sideset{}{'}\sum  \hat{\alpha}_{\v k}^{\dagger}  \hat{\alpha}_{\v k}$
commutes with both Hamiltonians (\ref{h2}) and (\ref{hi3}) at this point.
Therefore we can choose the basis in which $\hat{N}$ is diagonalized with given value $N$.
This gives the condition $\hat{N}_0+\sideset{}{'}\sum  \hat{\alpha}_{\v k}^{\dagger}  \hat{\alpha}_{\v k}=N$ which
allows us to eliminate the zero mode occupation number operator $\hat{N}_0$.
Using an expansion for the square root function and keeping up to the second order in the new set of
creation and annihilation operators, we obtain our approximated Hamiltonian as
\be \label{ht}
\hat{H}_{{\rm tot}}  =E_0 + \sideset{}{'}\sum (\epsilon _{k}+gn) \hat{\alpha}_{\v k}^{\dagger}\hat{\alpha}_{\v k}
+ \frac{gn}{2} \sideset{}{'}\sum ( \hat{\alpha}_{\v k}^{\dagger}\hat{\alpha}^{\dagger}_{-\v k} + {\rm h.c.})\\
+ \frac {n \lambda}{\sqrt{N}} \sideset{}{'}\sum (\tilde{\rho} _{\v k}(t) \hat{\alpha}_{\v k}^{\dagger} +{\rm h.c.}) .
\ee
Here $E_0 = g n N/2 -gn/2+ \lambda n$ is a constant term,
and we neglected terms of order of $N^{-1}$.

\subsection{Diagonalization}
Let us implement the Bogoliubov transformation to diagonalize the bilinear terms in the Hamiltonian (\ref{ht}).
The required canonical transformation $U=\exp (i \hat{G}_B)$ is generated by
\be \label{GB}
\hat{G}_B = \frac i2 \sideset{}{'}\sum \theta _{k}  \hat{\alpha}_{\v k}^{\dagger}  \hat{\alpha}_{-\v k}^{\dagger}+{\rm h.c.},
\ee
with
\be
 \theta _{k} = \tanh ^{-1}(\frac{gn}{\hbar \omega _{k} +\epsilon _{k}+gn}), \ \hbar \omega _{k}
 = \sqrt{\epsilon _{k}(\epsilon _{k}+2 gn)} .
\ee
Bogoliubov's excitation (the ``bogolon") is created and annihilated by the operators
\begin{align} \label{bog1}
\hat{b}_{\v k}^{\dagger}  &=
e^{i \hat{G}_B } \hat{\alpha}_{\v k} ^{\dagger}e^{- i \hat{G}_B }
= \hat{\alpha}_{\v k} ^{\dagger} \cosh \theta _{k} +\hat{\alpha}_{-\v k} \sinh \theta _{k}, \\
\hat{b}_{\v k}& = e^{i \hat{G}_B} \hat{\alpha}_{\v k} e^{- i \hat{G}_B}
=\hat{\alpha}_{\v k}  \cosh \theta _{k} + \hat{\alpha}_{-\v k}^{\dagger} \sinh \theta _{k}.\label{bog2}
\end{align}
These operators satisfy the same canonical commutation relations as before.
The new vacuum state of the Fock space ${\cal H}_B$ for the bogolon is defined by
\be
\hat{b}_{\v k}(t) | 0_B \rangle =0 \quad {\rm for\ all}\ \v k \neq 0.
\ee
This new vacuum contains infinitely many modes created by the original creation operators $\hat{\alpha}_{\v k}^{\dagger}$
\begin{align} \label{vacB}
| 0_B \rangle & = \exp (i \hat{G}_B) | 0 \rangle \\ \nonumber
&=\exp\{ \sideset{}{'} \sum [-\ln (\cosh \theta _{k}) +\hat{\alpha}_{\v k} ^{\dagger}\hat{\alpha} _{- \v k} ^{\dagger}
\tanh \theta _{k})]\} | 0 \rangle .
\end{align}
It is well known that two Fock spaces ${\cal H}_0$ and ${\cal H}_B$ are unitarily inequivalent to each other
because they possess infinite degrees of freedom.
The above formal expression (\ref{vacB}) loses its meaning in the thermodynamic limit.
Therefore we need to specify the vacuum state corresponding to each Fock space \cite{umezawa1, umezawa2}.
To see this point more clearly, we evaluate the inner product of two vacua $|0\rangle$ and $|0_B\rangle$:
\be
\langle 0|0_B\rangle=\exp[-\sideset{}{'}\sum\ln(\cosh\theta_k)]\\ \to\exp[-V \int \frac{d^3k}{(2\pi)^3}\;\ln(\cosh\theta_k)].
\ee
Obviously this results in $\langle 0|0_B\rangle=0$ for the $V\to\infty$ limit, meaning $|0_B\rangle$ cannot
be written as a superposition of vectors belonging to the Hilbert space ${\cal H}_0$.

After the Bogoliubov transformation generated by (\ref{GB}), the total Hamiltonian (\ref{ht}) becomes
\be \label{hb}
\hat{H}_B= E'_0
+\sideset{}{'} \sum \hbar \omega _{k} \hat{b}_{\v k}^{\dagger}  \hat{b}_{\v k}
+ \sideset{}{'} \sum \v (f_{\v k}(t) \hat{b}_{\v k}^{\dagger}+{\rm h.c.}\v ) .
\ee
Here $E'_0= E_0 + \sideset{}{'} \sum  (\hbar \omega _{k} -\epsilon _{k}-gn)/2$
is the ground state energy without impurities, and
\be \label{f}
f_{\v k}(t)= n \lambda \sqrt{\frac{\epsilon _{k}}{N \hbar \omega _{k}}}\;\tilde{\rho} _{\v k}(t) .
\ee
In the thermodynamic limit, the second term in $E'_0$ can be evaluated as follows.
\bal
\frac 12\sideset{}{'} \sum  &(\hbar \omega _{k} -\epsilon _{k}-gn)
=\frac V2 \int \frac{d^3k}{(2\pi)^3}\;(\hbar \omega _{k} -\epsilon _{k}-gn)\\ \label{diver}
&=-N\frac{Mng^2}{\hbar}\int \d3k \;\frac{1}{k^2}+\frac{64gN(n a_s)^{3/2}}{15\sqrt{\pi}}.
\end{align}
Obviously, the first term diverges in eq.~(\ref{diver}). This divergence can be
understood as an artifact of the approximation for the general two-body interaction $V(\v x-\v x')$
by the delta function $g \delta(\v x-\v x')$. As a consequence, the Fourier transform of
interaction is constant $g$ for all $\v k$. Therefore, all wave numbers including the ultraviolet region
contribute with the same weight. The standard way to deal with this divergence is
to renormalize the coupling constant $g$ by
\be
\frac{4\pi \hbar^2a_s}{M}=g-\frac{Mg^2}{\hbar}\int \d3k \;\frac{1}{k^2}.
\ee
This renormalization is done with the term $gnN/2$ in $E_0$,
and corresponds to including the second order Born approximation
for the scattering length $a_s$ \cite{fetter}.

The first order impurity effects in our model are described by the model equivalent to the generalized time
dependent van Hove model. We will explain later that the factor $1/\sqrt{N}$ due to BEC
introduces the significant difference and gives rise to the macroscopic degree of freedom.
In this sense it is different from the van Hove model.

The Heisenberg equations for the bogolon operators are
\begin{align} \label{he1}
i \hbar \pd _t  \hat{b}^{\dagger}_{\v k} &=[ \hat{b}^{\dagger}_{\v k}, \hat{H}_B]=- \hbar \omega _{k} \hat{b}^{\dagger}_{\v k} - f^*_{\v k}(t),\\
i \hbar \pd _t  \hat{b}_{\v k} &=[ \hat{b}_{\v k}, \hat{H}_B]= \hbar \omega _{k} \hat{b}_{\v k} + f_{\v k}(t). \label{he2}
\end{align}
We solve this differential equation with
the boundary condition such that the operators evolve without impurities asymptotically at
the remote past $t=- \infty$. The solution is
\begin{align} \label{solution1}
\hat{b}^{\dagger}_{\v k}(t)&=\hat{b}^{\in\dagger}_{\v k}(t)+\phi ^*_{\v k}(t),\\
\hat{b}_{\v k}(t)&=\hat{b}^{\in}_{\v k}(t)+\phi _{\v k}(t) .\label{solution2}
\end{align}
Here $\phi _{\v k}(t)$ is a $c$-number function :
\begin{align}  \label{phi}
\phi _{\v k}(t)  &=  - \frac{i}{\hbar} e^{-i \omega _{k} t} \int ^t _{-\infty} dt'\;f _{\v k}(t')e^{i \omega _{k} t'}\\
&= - \frac{in\lambda}{\hbar} \sqrt{\frac{\epsilon _{k}}{N \hbar \omega _{k}}}I_{\v k} (t)e^{-i \omega _{k} t}  ,
\end{align}
where the integral $I_{\v k}(t)$ is defined by
\be \label{integral}
I_{\v k}(t) =\int ^t _{-\infty} d t'\;\tilde{\rho}_{\v k} (t')e^{i\omega_k t'}.
\ee
Alternatively we can directly diagonalize the Hamiltonian $\hat{H}_B$ using the time dependent
unitary transformation $U_C(t)=\exp\v (i\hat{G}_C(t)\v )$ as will be shown below \cite{gross}.

In the Heisenberg picture the time evolution of operators $O(t)$ is governed by
the Heisenberg equation $i\hbar \dot{O}(t)= [O(t), H(t)]$.
(Explicit time dependence is not assumed for $\hat{O}(t)$.)
The unitary transformation $U(t)=\exp\v(iG(t)\v)$ for the operator $O(t)$ introduces a new
operator $\tilde{O}(t)=UOU^{\dagger}$.
Time evolution of the new transformed operator $\tilde{O}(t)$
is then given by the Heisenberg equation $i\hbar \dot{ \tilde{O}}(t)= [\tilde{O}(t), \tilde{H}(t)]$.
The transformed Hamiltonian is $\tilde{H}(t)=U(H-i\hbar U^{\dagger}\dot{U})U^{\dagger}$,
where the left hand side is expressed in terms of new transformed operators.
In our case we implement the time dependent unitary transformation $U_C(t)$ generated by
\begin{equation} \label{GC}
\hat{G}_C(t) = -i \sideset{}{'}\sum _{\v k} \phi _{\v k}(t)  \hat{b}_{\v k}^{\dagger} +{\rm h.c.}  
\end{equation} 
$\phi _{\v k}(t)$ is a c-number time dependent function which will be determined later. 
This transformation shifts $ \hat{b}_{\v k}^{\dagger}(t)$ and $ \hat{b}_{\v k}(t)$:
\begin{equation} \label{bin}
\begin{split}
\hat{b}_{\v k}^{\in\dagger} (t)& =
e^{i G_C }\hat{b}_{\v k}^{\dagger}(t) e^{- i G_C}
=\hat{b}_{\v k} ^{\dagger}(t)-  \phi _{\v k}^{*}(t), \\
\hat{b}^{\in}_{\v k} (t)& =
e^{i G_C }\hat{b}_{\v k}(t) e^{- i G_C}
= \hat{b}_{\v k}(t) -  \phi _{\v k}(t) .
\end{split}
\end{equation}
To obtain the transformed Hamiltonian in terms of $b_{\v k}^{\in\dagger}(t) $ and $b_{\v k}^{\in} (t)$,
we need to evaluate $K(t)= U_C^{\dagger}\dot{U}_C$.
The formula
\be
\exp(-iG)\frac{d}{dt} \exp(iG)\\=i\dot{G}+i^2[\dot{G},G]/2!+i^3[[\dot{G},G],G]/3!+\cdots,
\ee
after simple algebra yields,
\be 
K(t)=(i\omega_k\phi _{\v k}+\dot{\phi}_{\v k} )\hat{b}_{\v k}^{\dagger}+(i\omega_k \phi^*_{\v k}-\dot{\phi}_{\v k}^*)\hat{b}_{\v k}\\
+\frac 12 [(i\omega_k \phi_{\v k}-\dot{\phi}_{\v k}^*) \phi _{\v k}+(i\omega_k\phi _{\v k}+\dot{\phi}_{\v k} )\phi _{\v k}^*] .
\ee
The transformed Hamiltonian $\tilde{H}_B(t)$ is
\begin{align}
\tilde{H}_B(t)&=U_C(H_B(t)-i \hbar K(t))U_C^{\dagger} \\
&= \tilde{E}_0(t)+ \sideset{}{'} \sum \hbar \omega _{k} \hat{b}^{\in \dagger}_{\v k}(t) \hat{b}^{\in}_{\v k}(t) \equiv H^{\in}(t),
\end{align}
provided that the c-number function $\phi_{\v k} (t)$ satisfies a simple differential equation:
\be
i \hbar \dot{ \phi} _{\v k}(t) = \hbar \omega _{k}\phi _{\v k}(t) + f_{\v k}(t) .
\ee
The solution with the initial condition $\phi _{\v k}(t=-\infty)=0$ is given by (\ref{phi}).
In the diagonalized Hamiltonian we defined the new ground state energy $\tilde{E}_0(t)$ by
\be \label{newGE}
\tilde{E}_0(t)=E'_0+ \sideset{}{'} \sum {\rm Re}\v (f_{\v k}^*(t)\phi _{\v k}(t)\v ).
\ee

The asymptotic creation and annihilation operators denoted by $\hat{b}^{\in\dagger}_{\v k}(t)$ and
$\hat{b}^{\in}_{\v k}(t)$ respectively satisfy the same canonical commutation
relations, and time evolution is governed by the Heisenberg equation with the diagonalized Hamiltonian $\hat{H}^{\in}$,
i.e. $\hat{b}^{\in \dagger}_{\v k}(t)=e^{i\omega_k t}\hat{b}^{\in \dagger}_{\v k}$ and
 $\hat{b}^{\in}_{\v k}(t)=e^{-i\omega_k t}\hat{b}^{\in}_{\v k}$. Hence,
 \be \label{hin}
\hat{H}^{\in}(t)=\tilde{E}_0(t)+ \sideset{}{'} \sum \hbar \omega _{k} \hat{b}^{\in \dagger}_{\v k} \hat{b}^{\in}_{\v k}.
\ee

\begin{figure}[htbp]
   \centering
   \includegraphics[width=3.5in]{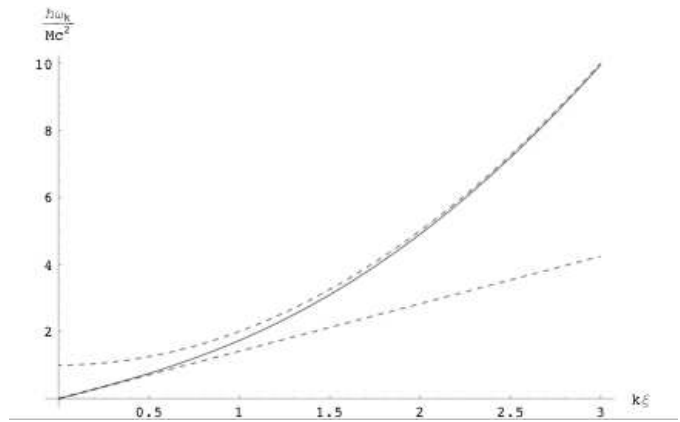}
   \caption{Bogoliubov's excitation spectrum $\hbar\omega_k=\sqrt{\epsilon_k(\epsilon_k+2gn)}$.
$\hbar\omega_k/(Mc^2)$ is plotted as a function of $\hbar k/(\sqrt2 Mc)$.  
Two asymptotic curves are also plotted. The dashed line represents $\hbar kc$ ($k\xi\ll 1$). 
The dashed curve represents $\epsilon_k +Mc^2$ ($k\xi\gg 1$). }
   \label{fig1}
\end{figure}

\begin{figure}[htbp]
   \centering
   \includegraphics[width=3.5in]{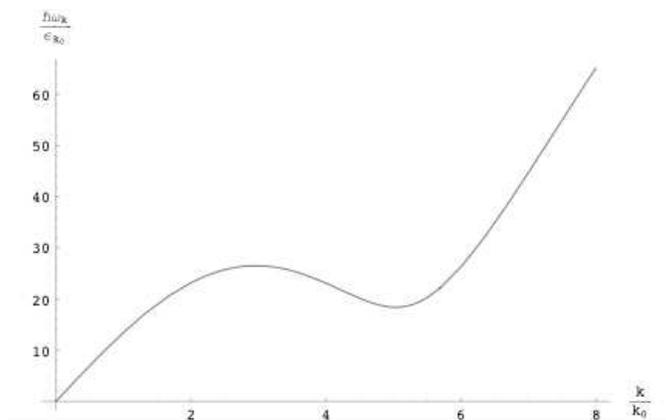}
   \caption{Bogoliubov's excitation spectrum $\hbar\omega_k=\sqrt{\epsilon_k(\epsilon_k+2n\tilde{V}(k))}$ for (\ref{bogdis2}).
   $\hbar \omega_k/\epsilon_{k_0}$ is plotted as a function of $k/k_0$ with $k_0=1/r_0$. }
   \label{fig2}
\end{figure}

The energy spectrum $\hbar \omega _{k}$ is that of the gapless excitations characterized
by $\omega _{k} \simeq kc$ for a small $k$, where $c=\sqrt{gn/M}$ is the speed of sound.
Fig.~\ref{fig1} shows the characteristics of Bogoliubov's spectrum.
We remark that the spectrum $\omega_k$ and the speed of sound $c$ are the same
as in the original Bogoliubov model without impurities.
Therefore the motion of impurities does not affect either $\omega_k$ nor $c$ in our approximation.
Another remark regarding Bogoliubov's spectrum is that the general two-body interaction $V (|\v x-\v x'|)$ instead of
the contact interaction $g \delta(\v x-\v x')$ produces different curve as shown in Fig.~\ref{fig2}.
This Fig.~\ref{fig2} is plotted for the finite height potential
\be \label{twobody}
V(|\v x-\v x'|)=
\begin{cases}
V_0 \ (|\v x|<r_0)\\
0\ \;(|\v x|>r_0) ,
\end{cases}
\ee
with $V_0$ and $r_0$ constants.
The contact interaction approximation is understood as the limit of this potential.
In the general two-body interaction case, the excitation spectrum is
\be
\hbar  \omega_k= \sqrt{\epsilon_k(\epsilon_k+2n\tilde{V}(k))},
\ee
where $\tilde{V}(k)$ is the Fourier transform of the potential
\be
\tilde{V}(k)=\int d^3x\; V(|\v x|)e^{-i\v k\cdot\v x}.
\ee
For the above particular choice of interaction (\ref{twobody}),
\be \label{bogdis2}
\hbar \omega_k=\sqrt{\epsilon_k^2
+\frac{2\pi n\hbar^2r_0V_0}{M}\left(\frac{\sin kr_0}{kr_0}-\cos kr_0\right)}.
\ee
This excitation spectrum has also phonon-like behavior for small $k$, i.e. $\hbar \omega_k\simeq \hbar k c'$
with $c'=\sqrt{2\pi n V_0r_0^3/(3M)}$. Interestingly by taking into account the finite distance interaction,
Bogoliubov's excitation contains the roton-like excitation spectrum as seen in Fig.~\ref{fig2} \cite{preparata}.

The vacuum state of the Fock space $\cal{H}_{\in}$ for $\hat{b}^{\in\dagger}_{\v k}$ and
$\hat{b}^{\in }_{\v k}$ is defined by
\be
\hat{b}^{\in }_{\v k} |0_{\in}\rangle =0 \quad {\rm for\ all}\ \v k \neq 0,
\ee
and this vacuum represents the homogeneous condensate.
The relation between two vacua of Fock spaces ${\cal H}_{B}$ and ${\cal H}_{\in}$ is formally expressed as
\begin{align}
 |0_{\in}\rangle &= \exp\v (i\hat{G}_C(t)\v ) |0_{B}\rangle \\
 &= \exp [ \sideset{}{'}\sum \v (-\frac 12  | \phi _{\v k}(t) |^2 + \phi _{\v k}(t) \hat{b}_{\v k}^{\dagger} \v ) ] | 0 _B \rangle .
\end{align}
The homogeneous condensate vacuum $ |0_{\in}\rangle$ is time independent
since the time dependence of the annihilation operator $\hat{b}^{\in }_{\v k}(t)$ is simply $\exp(-i\omega_k t)$.
Hence the Hilbert space of bogolons ${\cal H}_{B}[t]$ is a one parameter family of ${\cal H}_{\in}$ characterized by the time $t$.

The physical interpretation of the above result is quite simple.
The motion of impurities creates the time dependent coherent states.
The dressed Bogoliubov's excitations do not annihilate the homogeneous condensate vacuum:
\be \label{result}
\hat{b}_{\v k} (t) |0_{\in}\rangle =  \phi _{\v k}(t) |0_{\in}\rangle .
\ee

\section{Occupation Number and Energy Dissipation}
One of our main effects described by our model is the effect of
energy dissipation due to the motion of impurities in the
homogeneous BEC. From our model we find that even though the
systems is in the ground state where no elementary excitation is
present, there is energy transferred from the motion of impurities
to massive bosons. In other words, the motion of impurities will
dress bogolons in such a way that bogolons will not see the
homogeneous condensate as a vacuum. Since we do not take into
account any back reaction on the impurities, this transferred
energy is the energy in order for the impurities to move along
given trajectories. It means that we need to feed this amount of
energy to keep the impurities moving. More realistic model for the
dissipation effects in BEC should be done by considering quantum
impurities and taking into account the dynamics of them.

We first evaluate the occupation number for bogolons with respect to the homogeneous condensate.
This number counts the emitted bogolons accompanying with the motion of impurities in BEC,
\be \label{occu}
\tilde{n}_{\v k}(t) \equiv \langle 0_{\in}| \hat{b}^{\dagger}_{\v k} (t)\hat{b}_{\v k} (t) |0_{\in} \rangle
= |\phi _{\v k} (t)|^2 =\frac{n^2 \lambda ^2 \epsilon_k}{N \hbar^3 \omega_{k}}|I_{\v k}(t)|^2.
\ee
Fluctuation in the occupation number $\Delta \tilde{n}_{\v k}(t)$ is
\begin{align}
\Delta \tilde{n}_{\v k}(t) &= [\langle 0_{\in}| \v (\hat{b}^{\dagger}_{\v k} (t)\hat{b}_{\v k} (t)\v )^2 |0_{\in} \rangle 
-\v (\langle 0_{\in}| \hat{b}^{\dagger}_{\v k} (t)\hat{b}_{\v k} (t) |0_{\in} \rangle\v )^2]^{1/2}\\
&= |\phi _{\v k} (t)|.
\end{align}
We are interested in evaluating the occupation number for the $t\to \infty$ limit,
which covers the whole trajectories of impurity motion.

We next evaluate the expectation value for the number operator
$\hat{N}_{\v k} (t)= \hat{a}_{\v k}^{\dagger}(t)\hat{a}_{\v k}(t)=\hat{\alpha}_{\v k}^{\dagger}(t)\hat{\alpha}_{\v k}(t)$ 
($\v k\neq 0$) with respect to the homogeneous condensates $|0_{\in}\rangle$ as
\be
\langle 0_{\in}| \hat{N}_{\v k} (t) |0_{\in} \rangle=
\frac 12( \frac{\epsilon _{k} +gn}{\hbar \omega _{k}}-1)
+\frac 12( \frac{\epsilon _{k}}{\hbar \omega _{k}}+1)|\phi _{\v k}(t)|^2\\
+\frac 12( \frac{\epsilon _{k}}{\hbar \omega _{k}}-1)|\phi _{-\v k}(t)|^2
+\frac{gn}{2\hbar \omega _{k}}|\phi^*_{\v k}(t)-\phi_{-\v k}(t)|^2.
\ee
The first term is the well known term in the original Bogoliubov argument, and the rest are
contributions due to the impurities. The depletion of the condensate $d(t)$ due to the quantum fluctuation can be
evaluated by summing over all modes and divided by the total particle number $N$:
\be
d(t)=   \frac 83 \sqrt{\frac{n a_s^3}{\pi}}
+ \frac 1N  \sideset{}{'} \sum (\frac{\epsilon _{k}}{\hbar \omega _{k}}|\phi _{\v k}(t)|^2\\
+\frac{gn}{2\hbar \omega _{k}}|\phi^*_{\v k}(t)-\phi_{-\v k}(t)|^2) .
\ee
Since $|\phi_{\v k}|^2$ has an additional factor $1/N$, the motion of impurities does not contribute
to the depletion of the homogeneous condensate in the thermodynamic limit.
Therefore, in the large $N$ limit, the homogeneous condensate is stable against
a small number of external impurities, i.e. when the
number of impurities is much less than the total number of bosons.
In contrast, real experiments have always the finite number
of particles and a finite size. Therefore, there are effects on the depletion of condensates due
to the motion of the impurities.

The dissipated energy ${\cal E}_{\v k} (t)$ for a given mode $\v k$ is obtained by multiplying $\tilde{n}_{\v k}(t)$
by the excitation energy $\hbar \omega _k$, i.e. ${\cal E}_{\v k} (t)=\hbar \omega _k \tilde{n}_{\v k}(t)$.
Therefore the total dissipated energy ${\cal E} (t)$ is given by summing over all modes,
which is expressed in the thermodynamic limit as
\be \label{diss}
{\cal E} (t)= \sideset{}{'} \sum {\cal E}_{\v k} (t) \to
\frac{n \lambda ^2 }{ \hbar^2} \int \d3k \epsilon_k |I_{\v k}(t)|^2.
\ee
In particular for the $t \to \infty$ limit, taking into account the complete trajectory of impurity motion,
we obtain rather a simple expression,
\begin{align}
{\cal E}_{{\rm tot}}&=\lim _{t \to \infty} {\cal E} (t)\\
&= \frac{n \lambda ^2 }{ 2M} \int \d3k \int d\omega
\;\delta ( \omega- \omega_k)k^2 |\tilde{\rho}_c(\v k, \omega)|^2.
\end{align}
Here $\tilde{\rho}_c(\v k, \omega)$ is the Fourier transform of impurity distribution $\rho_c(\v x,t)$
with respect to space and time coordinates :
\be \label{rhof}
\tilde{\rho}_c(\v k, \omega) =
\int dt \int d^3 x\;\rho_c(\v x ,t)e^{-i \v k \cdot \v x+i\omega t}.
\ee

\section{Emergent Macroscopic Fields}
The coherent states play an essential role in representing classical fields
from quantum systems in various areas of physics. In quantum electrodynamics
for example, the classical Maxwell equation can be derived as the Ehrenfest equation
with respect to the coherent state.
In our model one of the main results eq.~(\ref{result}) is best understood form this point of view.
To this end we express operators in terms of incoming field creation and annihilation operators
$\hat{b}^{\in\dagger}_{\v k}$ and $\hat{b}^{\in }_{\v k}$.

We first express the field operator $\hat{\psi}(\v x, t)$ in terms of incoming fields.
For this purpose we need to know the time dependence of operator $\hat{\beta}_0(t)$ defined by eq.~(\ref{beta}).
Using the Hamiltonian (\ref{h2}, \ref{hi3}), it is easy to show that the Heisenberg equation for this operator is
$i\hbar \pd_t\hat{\beta}_0(t)=gn\hat{\beta}_0(t)$ within our approximations, and hence
\be
\hat{\beta}_0(t) = e^{-i gnt/\hbar}\hat{\beta}^{\in}_0=e^{-i Mc^2t/\hbar}\hat{\beta}^{\in}_0.
\ee
Using the Bogoliubov transformation (\ref{bog2}) and the solution (\ref{solution2}),
\begin{align} \label{psi01}
\hat{\psi}(\v x, t)&=\hat{\beta}^{\in}_0 e^{-i Mc^2t/\hbar}
[\sqrt{n}+\frac{1}{\sqrt{V}}\sideset{}{'}\sum\v (\cosh \theta_k\;\hat{b}_{\v k}(t)e^{i\v k\cdot \v x}
-\sinh \theta_k\;\hat{b}^{\dagger}_{\v k}(t)e^{-i\v k\cdot \v x}\v ) ]\\
&=\hat{\beta}^{\in}_0 e^{-i Mc^2t/\hbar} \v (\psi_c(\v x, t)+\hat{\psi}_q(\v x, t)\v ).
\end{align}
Here the c-number field $\psi_c(\v x, t)$ and q-number field $\hat{\psi}_q(\v x, t)$ are defined by
\begin{align}
\psi_c(\v x, t)&=\sqrt{n}[1-
\frac{\lambda}{\hbar V}\sideset{}{'}\sum \frac{i}{\sqrt 2 \hbar \omega_k} \v (\hbar \omega_k I_{\v k}(t)e^{i\v k\cdot \v x-i\omega_kt}
-\epsilon_k I^*_{\v k}(t)e^{-i\v k\cdot \v x+i\omega_kt}\v ) ], \\ 
\hat{\psi}_q(\v x, t)&= \frac{1}{\sqrt{V}}\sideset{}{'}\sum\frac{1}{\sqrt{2\hbar \omega_k}}\v (\sqrt{gn+\epsilon_k+\hbar \omega_k}\hat{b}^{\in}_{\v k}e^{i\v k\cdot \v x-i\omega_k t}
-\sqrt{gn+\epsilon_k-\hbar \omega_k}\hat{b}^{\in\dagger}_{\v k}e^{-i\v k\cdot \v x}\v +i\omega_k t),
\end{align}
with $I_{\v k}(t)$ eq.~(\ref{integral}).
A similar expression for $\hat{\psi}^{\dagger}(\v x, t)$ can be obtained by taking the hermite conjugate of the above result.

We next evaluate the density operator $\hat{\rho}(\v x, t)$ and the current density operator $\hat{\v \jmath}(\v x, t)$ defined by
\begin{align}
\hat{\rho}(\v x, t)&=  \hat{\psi}^{\dagger} (\v{ x},t) \hat{\psi} (\v{ x},t), \\
\hat{\v \jmath}(\v x, t)&=
\frac{\hbar}{2Mi}[\hat{\psi}^{\dagger} (\v{ x},t)  \v (\v \nabla \hat{\psi} (\v{ x},t)\v )
- \v (\v \nabla\hat{\psi}^{\dagger} (\v{ x},t) \v )\hat{\psi} (\v{ x},t)] .
\end{align}
We express these operators in terms of incoming fields in the same manner as before,
\begin{align}\label{rhohat}
\hat{\rho}(\v x, t)&= n +\frac{1}{V} \sideset{}{'}\sum \sqrt{\frac{N \epsilon_k}{\hbar \omega_k}} \v (\phi^*_{\v k}(t)e^{-i\v k \cdot \v x}+{\rm c.c.}\v )
+\frac{1}{\sqrt{V}} \sideset{}{'}\sum \sqrt{\frac{n \epsilon_k}{\hbar \omega_k}} \v (\hat{b}^{\in\dagger}_{\v k}(t)e^{-i\v k \cdot \v x}+{\rm h.c.}\v ),\\ \label{jhat}
\hat{\v \jmath}(\v x, t)&=
\frac{1}{V} \sideset{}{'}\sum\frac{\hbar \v k}{2 M} \sqrt{\frac{N\hbar \omega_k}{\epsilon_k}}\v (\phi^*_{\v k}(t)e^{-i\v k \cdot \v x}+{\rm c.c.}\v )
+\frac{1}{\sqrt{V}} \sideset{}{'}\sum\frac{\hbar \v k}{2 M}
\sqrt{\frac{n\hbar \omega_k}{\epsilon_k}}\v (\hat{b}^{\in\dagger}_{\v k}(t)e^{-i\v k \cdot \v x}+{\rm h.c.}\v ) .
\end{align}
These equations satisfy the continuity equation :
\be
\pd _t \hat{\rho}(\v x, t)+\v \nabla \cdot \hat{\v \jmath}(\v x, t) =0,
\ee
as required by the self-consistency of our approximations.
In the thermodynamic limit, these operators have non-vanishing expectation values with respect to the homogeneous condensate $| 0_{\in} \rangle$. We now define the induced density due to the motion of an impurity by
\be
\bar{\rho}(\v x, t)=\langle 0_{\in}| \hat{\rho}(\v x,t) |0_{\in} \rangle-\langle 0_{\in}| \hat{\rho}(\v x,t) |0_{\in} \rangle |_{\lambda=0}.
\ee
In the homogeneous BEC case, this induced density accounts for the deviation from the constant density $n$.
The same expression for the induced current density is defined. Then eqs. (\ref{rhohat},\ref{jhat}) yields
\begin{align}
\bar{\rho}(\v x, t) &=\kappa c^2\int \d3k \frac{k^2}{2 \omega_k}(i I^*_{\v k}(t)e^{-i\v k \cdot \v x+i \omega_k t} +{\rm c.c.}),\\
\bar{\v \jmath}(\v x, t) &= \kappa c^2 \int \d3k \frac{\v k}{2} (i I^*_{\v k}(t)e^{-i\v k \cdot \v x+i \omega_k t} +{\rm c.c.}),
\end{align}
where we denote the ratio of two coupling constants $g$ and $\lambda$ by $\kappa=\lambda/g$.
These induced density and current density also satisfy the continuity equation :
\be
\pd _t \bar{\rho}(\v x, t)+\v \nabla \cdot \bar{\v \jmath}(\v x, t) =0.
\ee
Although the Bogoliubov excitation frequency $\omega_k$ contains $\hbar$,
the cancelation of $\hbar$ in the above expressions should be noticed.

It is also convenient to obtain the Fourier transform expression in terms of $\tilde{\rho}_c(\v k, \omega)$ (\ref{rhof}).
Using the formula
\be
\lim_{T\to\infty} \int ^{T}_0 dt\;e^{iwt}=\int_0^{\infty}dt\;e^{(iw-\epsilon)t}=\frac{i}{w+i\epsilon},
\ee
with $\epsilon$ a positive small
number which goes to zero at the end of calculations, $\bar{\rho}(\v x, t)$ and $\bar{\v \jmath}(\v x, t)$ become
\begin{align} \label{indden}
\bar{\rho}(\v x, t) &=\kappa c^2 \int \d3k \int \frac{d \omega}{2\pi} \frac{k^2}{2\omega_k}\tilde{\rho}_c(\v k, \omega)
(\frac{1}{\omega-\omega_k+i\epsilon}-\frac{1}{\omega+\omega_k+i\epsilon})e^{i\v k\cdot \v x-i\omega t}, \\
\bar{\v \jmath}(\v x, t)&=\kappa c^2 \int \d3k \int \frac{d \omega}{2\pi} \frac{\v k}{2} \tilde{\rho}_c(\v k, \omega)
(\frac{1}{\omega-\omega_k+i\epsilon}+\frac{1}{\omega+\omega_k+i\epsilon})e^{i\v k\cdot \v x-i\omega t} . \label{indcur}
\end{align}
Fourier transforms of $\bar{\rho}(\v x, t)$ and $\bar{\v \jmath}(\v x, t)$ are written
using the Fourier transform of the retarded Green function, ${\rm D}_{{\rm ret}}(\omega, \omega_k)=1/(\omega^2-\omega_k^2+i{\rm sgn}(\omega)\epsilon)$,
\begin{align}
\bar{\rho}(\v k, \omega) &=
\kappa c^2k^2 \tilde{\rho}_c(\v k, \omega){\rm D}_{{\rm ret}} (\omega, \omega_k),\\
\bar{\v \jmath}(\v k,\omega)&=
\kappa c^2\v k \omega \tilde{\rho}_c(\v k, \omega){\rm D}_{{\rm ret}}(\omega, \omega_k)  .
\end{align}
The induced density satisfies the following differential equation:
\be \label{waveeq}
(\pd_t^2-c^2\v{\triangle}+ c^2\xi^2\v{\triangle}^2)\bar{\rho}(\v x, t)=\kappa c^2\v{ \triangle} \rho_c(\v x, t),
\ee
where $\v{\triangle}=\v{ \nabla} ^2$ is the Laplacian and
$\xi=\hbar/(2 Mc)$ is the coherence length of the homogeneous condensate.
We can also obtain the differential equation for the induced current $\bar{\v{\jmath}}(\v x,t)$ in the same manner.
Assuming the distribution of impurities $\rho_c(\v x,t)$ satisfies the continuity equation
$\pd_t \rho_c(\v x,t)+\v{\nabla}\cdot \v{\jmath}_c(\v x,t)=0$, e.g.
for a point-like impurity $\rho_c =\delta(\v x-\v\zeta (t))$ and $\v \jmath_c=\dot{\v \zeta}(t)\delta(\v x-\v\zeta (t))$, we have
\begin{align} \label{waveeq3}
(\pd_t^2-c^2\v{\triangle}+ c^2\xi^2\v{\triangle}^2)\bar{\v{\jmath}}(\v x, t)
&=-\kappa c^2\v{\nabla}\pd_t\rho_c(\v x, t)\\&=\kappa c^2\v{\nabla}\v (\v{\nabla}\cdot \v{\jmath}_c(\v x,t)\v ).
\end{align}

Lastly we evaluate the induced stress energy momentum complex $\bar{T}^{\mu}_{\ (\nu)}(\v x,t)$ ($\mu, \nu=0,1,2,3$)
defined by
\be
\bar{T}^{\mu}_{\ (\nu)}(\v x, t)=\langle 0_{\in}| \hat{T}^{\mu}_{\ (\nu)}(\v x,t) |0_{\in} \rangle
-\langle 0_{\in}| \hat{T}^{\mu}_{\ (\nu)}(\v x,t) |0_{\in} \rangle |_{\lambda=0}.
\ee

The stress energy momentum complex $\hat{T}^{\mu}_{\ (\nu)}(\v x,t)$ of the Bogoliubov model is obtained from
the Lagrangian density :
\be
{\cal L}=\frac{i\hbar}{2}(\hat{\psi}^{\dagger}   \pd_t \hat{\psi}  -\pd_t \hat{\psi}^{\dagger}  \hat{\psi})
-\frac{\hbar^2}{2M}\pd_i \hat{\psi}^{\dagger}   \pd_i \hat{\psi}
-\frac g2 \hat{\psi}^{\dagger}   \hat{\psi}^{\dagger}   \hat{\psi}   \hat{\psi},
\ee
where the space-time coordinate dependence $(\v x,t)$ is for simplicity omitted for $\hat{\psi}^{\dagger}(\v x,t)$ and $\hat{\psi}(\v x,t)$.
When the action $\int d^3x \int dt {\cal L}$ is invariant under an infinitesimal space-time translation,
there exists the stress energy momentum complex of four conserved N\"oether currents \cite{takahashi3,takahashi86}.
These four conserved currents are defined by
\be
\hat{T}^{\mu}_{\ (\nu)}=\frac{\pd {\cal L}}{\pd (\pd _{\mu}\hat{\psi})}\pd_{\nu}\hat{\psi}
+\pd_{\nu} \hat{\psi}^{\dagger}\frac{\pd {\cal L}}{\pd (\pd _{\mu} \hat{\psi}^{\dagger})}-\delta^{\mu}_{\ \nu}{\cal L},
\ee
where $\pd_0=\pd_t$ and summation of repeated indices is understood.
For the Bogoliubov model,
\begin{align}\label{tensor}
\begin{split}
\hat{T}^{0}_{\ (0)}&=\frac{\hbar^2}{2M}\pd_i \hat{\psi}^{\dagger}\pd_i \hat{\psi}
+\frac g2 \hat{\psi}^{\dagger}\hat{\psi}^{\dagger}\hat{\psi}\hat{\psi} ,\\
\hat{T}^{i}_{\ (0)}&=-\frac{\hbar^2}{2M}(\pd_t \hat{\psi}^{\dagger}\pd_i \hat{\psi}+\pd_i \hat{\psi}^{\dagger}\pd_t \hat{\psi}),\\
\hat{T}^{0}_{\ (i)}&=\frac{i\hbar}{2}(\hat{\psi}^{\dagger}\pd_i \hat{\psi}  -\pd_i \hat{\psi}^{\dagger}  \hat{\psi}),\\
\hat{T}^{i}_{\ (j)} &=-\frac{\hbar^2}{2M}(\pd_i \hat{\psi}^{\dagger}  \pd_j \hat{\psi}  +\pd_j \hat{\psi}^{\dagger}  \pd_i \hat{\psi}  )\\
&\hspace{2cm}+\delta_{ij}[-\frac{i\hbar}{2}(\hat{\psi}^{\dagger}\pd_t \hat{\psi}-\pd_t \hat{\psi}^{\dagger}\hat{\psi})
+ \frac{\hbar^2}{2M}\pd_k \hat{\psi}^{\dagger}\pd_k \hat{\psi}  +\frac g2 \hat{\psi}^{\dagger}\hat{\psi}^{\dagger}\hat{\psi}   \hat{\psi} ].
 \end{split}
\end{align}
These four N\"oether currents satisfy the continuity equation $\pd_{\mu}\hat{T}^{\mu}_{\ (\nu)}=0$ ($\nu=0,1,2,3$),
which results in conservation of the energy ($\nu=0$) and the momentum ($\nu=1,2,3$).
For the general two-body interaction case, the stress energy momentum complex does not conserve locally, but
it does conserve globally.

Substituting eq.~(\ref{psi01}) and its hermite conjugate into eqs.~(\ref{tensor}), it is easy to show that the Fourier transform of
the induced stress energy momentum complex,
\be
\bar{T}^{\mu}_{\ (\nu)}(\v k,\omega)=\int d^3x \int dt \;\bar{T}^{\mu}_{\ (\nu)}(\v x,t)e^{-i\v k \cdot \v x+i\omega t},
\ee
is
\begin{align}
\bar{T}^{0}_{\ (0)}(\v k, \omega)&=\kappa Mc^4k^2 \tilde{\rho}_c(\v k, \omega){\rm D}_{{\rm ret}} (\omega, \omega_k),\\
\bar{T}^{i}_{\ (0)} (\v k, \omega)&=\kappa Mc^4 k_i \omega \tilde{\rho}_c(\v k, \omega){\rm D}_{{\rm ret}} (\omega, \omega_k),\\
\bar{T}^{0}_{\ (i)} (\v k, \omega)&=-\kappa Mc^2 k_i \omega \tilde{\rho}_c(\v k, \omega){\rm D}_{{\rm ret}} (\omega, \omega_k),\\
\bar{T}^{i}_{\ (j)} (\v k, \omega)&=-\delta_{ij}\kappa Mc^2\omega^2 \tilde{\rho}_c(\v k, \omega){\rm D}_{{\rm ret}} (\omega, \omega_k).
\end{align}

From the above result, we see that it is natural to define the effective Minkowski metric tensor
\be \label{eta1}
\eta _{\mu \nu}={\rm diag}(c^2, -1,-1,-1),
\ee
with $c$ the speed of sound, and its inverse
\be \label{eta2}
\eta^{\mu \nu}={\rm diag}(1/c^2, -1,-1,-1).
\ee
Then we raise and lower the indices to get the symmetric stress energy momentum complex defined by
$\bar{T}^{\mu \nu}(\v k,\omega)=\eta^{\mu\nu}\bar{T}^{\mu}_{\ (\nu)}(\v k,\omega)$ :
\begin{align}
\bar{T}^{00}(\v k, \omega)&=\kappa Mc^2k^2 \tilde{\rho}_c(\v k, \omega){\rm D}_{{\rm ret}} (\omega, \omega_k),\\
\bar{T}^{i0} (\v k, \omega)&=\kappa Mc^2 k_i \omega \tilde{\rho}_c(\v k, \omega){\rm D}_{{\rm ret}} (\omega, \omega_k),\\
\bar{T}^{0i} (\v k, \omega)&=\kappa Mc^2 k_i \omega \tilde{\rho}_c(\v k, \omega){\rm D}_{{\rm ret}} (\omega, \omega_k),\\
\bar{T}^{ij} (\v k, \omega)&=\delta_{ij}\kappa Mc^2\omega^2 \tilde{\rho}_c(\v k, \omega){\rm D}_{{\rm ret}} (\omega, \omega_k) .
\end{align}
This induced stress energy momentum complex is conserved locally in our approximation
\be
\pd_{\mu} \bar{T}^{\mu \nu}(\v x, t)=0\;(\nu=0\sim 3).
\ee

Several remarks are in order.
We first observe that for very large length scale, much
greater than $\xi$, we can approximate eq.~(\ref{waveeq}) as
\be \label{waveeq2}
\eta^{\mu \nu} \pd_{\mu}\pd_{\nu}\bar{\rho}(\v x, t)=\kappa \v{ \triangle} \rho_c(\v x, t).
\ee
Thus bogolons propagate following the ordinary sound wave equation in barotropic fluids with the speed of sound $c$
regardless the details of the motion of impurities.
We emphasize the non-triviality of this result. 
Although the homogenous BEC maintains its condensed state, where the depletion of condensate 
is very small in the large $N$ limit, 
there arises macroscopic change in its density 
$\langle 0_{\in}| \hat{\rho}(\v x,t)|0_{\in}\rangle=n+\bar{\rho}(\v x,t)$.
Especially, any small disturbance in the homogeneous BEC at some space-time point will
produce the propagation of bogolons inside the space-time light cone
as represented by the wave equation (\ref{waveeq2}).
In this large length scale, the left hand side of (\ref{waveeq2}) respects the Lorentz symmetry rather than
the original Galilean symmetry. This effective symmetry becomes more exact as we go to a larger length scale.
However, this classical behavior will not hold when we reach the high energy scale or the
very small length scale $\xi$, where we cannot neglect the terms depending on the coherence length $\xi$.
This situation is familiar: quantum hydrodynamics breaks down at certain length scales.
There, the impurities start to see the atoms rather than phonon-like low energy excitations.

From the induced stress energy momentum complex, we have $\bar{T}^{0}_{\ (0)}=Mc^2\bar{\rho}$, i.e.
the deviation of the energy density of this fluid is equal to $Mc^2\times$(the induced density). We also have
the relation $\bar{T}^{i}_{\ (0)}=\bar{T}^{0i}c^2$, in other words,
(energy density flux) = (momentum density)$\times c^2$. The famous Einstein relation holds true,
with $c$ the speed of sound.
Lastly, we point out that the induced stress energy momentum complex $\bar{T}^{\mu\nu}(\v x,t)$
is completely symmetric, i.e. $\bar{T}^{\mu\nu}(\v x,t)=\bar{T}^{\nu\mu}(\v x,t)$.
Here we used the inverse of the effective Minkowski metric tensor $\eta^{\mu\nu}$ to
raise the index: $\bar{T}^{\mu \nu}(\v x, t)=\eta^{\mu\nu}\bar{T}^{\mu}_{\ (\nu)}(\v x,t)$.
This is a non-trivial result.
This suggests that relativity could indeed emerge as the consequence of the existence of an underlying constituent.

\section{Examples}
We will consider several examples to illustrate our model and its general solution.
In order to discuss the analogy to special relativity,
we use the relativistic form of trajectories. 
(In order to discuss realistic experimental settings, we need to use 
the non-relativistic trajectories instead. 
Moreover, we should use finite times, volumes, and trajectories conforming to the experimental situations.) 
The speed of sound plays the role of the speed of light. 
We are interested in the occupation number (\ref{occu}) and
the total dissipated energy (\ref{diss}) for the $t\to \infty$ limit.
Therefore our main concern is to evaluate the integral (\ref{integral}) for
$\rho_c (\v x, t)=\delta(\v x- \v \zeta(t))$, with $\v \zeta (t)$ a trajectory of point-like impurity.
\begin{align}
I_{\v k}(\infty) &=\int dt \int d^3 x \;\rho_c(\v x ,t)e^{-i \v k \cdot \v x+i\omega_k t} \\
&=\int _{-\infty}^{\infty}dt \;e^{i\omega_k t-i\v k \cdot \v \zeta (t)}.   \label{int}
\end{align}

\subsection{Method of Stationary Phase and Landau's Criterion}
Since it is impossible to carry out the above integral (\ref{int}) in general,
let us first examine it using the method of stationary phase.

We are interested in the behavior of this integral for high momenta
where the condition $\omega_k \gg kc$ holds.
Then the integral oscillates rapidly, so that we can approximate
it using the method of stationary phase.
Rewrite the integral as
\be
I_{\v k}(\infty) =\int dt \;e^{i\omega_k g(t)},
\ee
where $g(t)=t-\v k \cdot \v{\zeta}(t)/\omega_k=t-\hat{\v k} \cdot \v{\zeta}(t)/(c\sqrt{1+(k\xi)^2})$,
with $\hat{\v k}=\v k/k$ a unit vector in the direction of $\v k$.
The stationary phase point is obtained by
\be
g'(\tau)=0 \leftrightarrow \hat{\v k} \cdot \dot{\v{\zeta}}(\tau)/c=\sqrt{1+(k\xi)^2} \ (>1), \label{station}
\ee
and the integral is approximated as
\begin{align}
I_{\v k}(\infty)& \simeq \sum_{j} e^{i\omega_k \tau_j-i\v k \cdot \v \zeta (\tau_j)} \int dt \;e^{-i\v k\cdot \ddot{\v \zeta}(\tau_j)(t-\tau_j)^2/2} \\
&= \sum_{j} e^{i\omega_k \tau_j-i\v k \cdot \v \zeta (\tau_j)} \sqrt{\frac{2\pi}{i\v k\cdot \ddot{\v \zeta}(\tau_j)}}.\label{stationformula}
\end{align}
Here the summation should be taken over the stationary phase points obtained by (\ref{station}), and
$\v k\cdot \ddot{\v \zeta}(\tau_j)\neq0$ is assumed. Otherwise the integral is approximated as zero.
In order to give rise to the manifest energy dissipations,
the argument of the exponent (\ref{stationformula}) has to be imaginary.
However as long as the velocity of the impurity is less than the speed of sound $c$,
no real $\tau$ exists satisfying (\ref{station}). This means any subluminal motion below $c$
will produce exponential suppression factors resulting in much smaller dissipation effects.
(In addition we need a condition ${\rm Im} \v (g(\tau_j) \v )>0$, otherwise the integral grows exponentially.)
This general statement is equivalent to Landau's criterion \cite{landau}.
Indeed, if we start from the general two-body interactions $V(\v x-\v x')$ rather than the contact interaction
as mentioned before, the above stationary phase point condition becomes
\be
\v k\cdot\!\dot{\v \zeta}(t)=\omega(k).
\ee
Here $\omega(k)= \sqrt{\epsilon_k(\epsilon_k+2n\tilde{V}(k))}/\hbar$ is the dispersion relation for the general two-body interaction.
Thus we obtain the critical velocity $v_{c}$ as
\be \label{criticalvel}
v_c=\min_{\v k}\left(\frac{\omega(k)}{|\v k|}\right).
\ee
The stationary phase points are always complex for $|\dot{\v \zeta}(t)|<v_c$.
The corresponding statement is that there is always an exponential suppression when the
speed of the impurity is smaller than the critical velocity $|\dot{\v \zeta}(t)|<v_c$.
Therefore, we derive Landau's criterion from the first principle based on our simple model.
We will give a more quantitative argument by considering the motion of an impurity with a constant velocity later.
As is clear from our discussion, an impurity moving slightly slower than the critical velocity can
give rise to energy dissipation. This cannot be explained from Landau's argument based
on the kinematical consideration.

\subsection{Two Stationary Impurities}
As a special case, we first consider two point-like impurities sitting in a BEC.
For simplicity we assume the impurities have the same mass, i.e. the coupling constant $\lambda$ is the same.
The distribution is $\rho_c (\v x)=(\delta(\v x- \v x_1)+\delta(\v x- \v x_2))/2$ where the factor
$1/2$ is a convention to make $\int d^3x \;\rho_c (\v x, t)=1$, and $\v x_1$, $\v x_2$ are constant vectors.
It is clear that this special case does not give rise to energy dissipation in the $t\to \infty$ limit, since
$I_{\v k}(\infty)\propto\delta(\omega_k)$, and $\omega_k=0$ if and only if $k=0$.
However, the ground state energy
\be
\tilde{E}_0(t)=E'_0+ \sideset{}{'} \sum {\rm Re}\v (f_{\v k}^*(t)\phi _{\v k}(t)\v ),
\ee
depends on the distance between the two impurities $R=|\v x_1-\v x_2|$.

The limit $t\to \infty$ should be taken carefully in this case, since the Fourier transform of $\rho_c (\v x)$
does not depend on time, i.e. $f_{\v k}$ is time independent.
\begin{align}
\nonumber{\rm Re}\v ( &f^*_{\v k}\phi_{\v k}(t) \v ) \\
&= {\rm Re}\left( \frac{-in^2\lambda^2\epsilon_k|\tilde{\rho}_{\v k}|^2e^{-i\omega_k t}}{N\hbar^2\omega_k}
\!\lim_{T\to\infty} \int ^t_{-T}dt' \;e^{i\omega_k t'}\right) \\
 &= \frac{-n^2\lambda^2\epsilon_k|\tilde{\rho}_{\v k}|^2}{N\hbar^2\omega_k}
\!\lim_{T\to\infty} \frac{1-\cos\omega_k(t+T)}{\omega_k} \\
&= \frac{-n^2\lambda^2\epsilon_k|\tilde{\rho}_{\v k}|^2}{N\hbar^2\omega_k^2} ,
\end{align}
where we use $\lim_{T\to\infty}(1-\cos wT)/w={\cal P}(1/w)$ with ${\cal P}$ the principal value distribution,
and ${\cal P}(1/w)=1/w$ for $w\neq0$. The above limit holds at any time $t$ except $t=-\infty$.
Hence the ground state energy is a constant of time, and in the thermodynamic limit,
\begin{align}
\tilde{E}_0&=E'_0-\frac{n\lambda^2}{2\hbar^2} \int \d3k \frac{\epsilon_k}{\omega_k^2}(1+\cos\v k\cdot\!\v R)\\
&=E'_0-\frac{Mn\lambda^2}{\hbar}\int \d3k \frac{1}{k^2}
+\frac{gn^2\lambda^2}{\hbar^2}\int \d3k \frac{1}{\omega_k^2}
-\frac{n\lambda^2}{2} \int \d3k \frac{\cos(\v k\cdot\!\v R)}{\epsilon_k+2gn}\\
&=E'_0+\frac{n\lambda^2M^2c}{2\pi \hbar^3}-\frac{n\lambda^2M}{4\pi \hbar^2}\;\frac{\exp(-R/\xi)}{R}. \label{yukawa}
\end{align}
The divergent integral, the second term in the second line, should
be considered as a term arising from the fact that we approximate
the interaction Hamiltonian (\ref{hi}) by taking only contact
interaction with the s-wave scattering length $b_s$. We have
already encountered the same divergent integral when we calculated
the ground state energy $E'_0$ without impurities by renormalizing
the coupling constant $g$. The same procedure is applied for the
coupling constant $\lambda$ which appears in $E'_0$.

The third term in eq.~(\ref{yukawa}) has the same form as the Yukawa potential
and gives rise to an attractive force between two stationary impurities.
In real experimental situations however, the distance between two impurities is much larger than the coherence length,
then attractive force decays with the exponential factor.

On the other hand, if the mass $M$ of boson happens to be very small such that the
coherence length has the same order as the size of the system $L=(V)^{1/3}$,
then the effective potential $V_{{\rm eff}}$ for the impurities can be approximated as
\be \label{newton}
V_{{\rm eff}}\simeq -\frac{n\lambda^2M}{4\pi \hbar^2}\;\frac 1R .
\ee
Thus the homogeneous condensate produces a long-range attractive force between impurities.
This is the same as the Newtonian gravity regardless of details of the impurities.
The strength of this attractive force depends only through the combination $n\lambda^2M/(4\pi \hbar^2)$.
Expressing the coupling constant $\lambda$ in terms of the s-wave scattering length $b_s$,
the coefficient of (\ref{newton}) is approximated $n\lambda^2M/(4\pi \hbar^2)\simeq n\pi b_s^2\hbar^2/M$.
In this scenario, the impurities with a longer s-wave scattering length get attracted stronger.

We remark that this effective attractive force was also obtained
by others by studying a mixture of bosons and fermions
\cite{meystre,stoof, pethick}. There they found that a bosonic
background produces an attractive Yukawa type force between
two-component fermions.

\subsection{Constant Velocity}
As the simplest example of the motion of an impurity in the homogenous BEC, let us consider
a linear motion of one impurity with a constant velocity $\v v$.
The trajectory is $\v \zeta(t)=\v v t$, then the integral (\ref{int}) is
\be
I_{\v k}(\infty)=2\pi\;\delta(\omega_k-\v k\cdot\v v).
\ee
To obtain the total dissipated energy in this case, one needs to give a little attention to the limit $t\to\infty$
since we cannot take a square $I_{\v k}(\infty)$ directly. To this end we first integrate for a finite
time interval $T$, and then we take the limit $T\to\infty$ at the end of calculations.
The dissipated energy becomes proportional to $T$ for large $T$ with the use of
the formula $(1-\cos wT)/w^2\simeq \pi T \delta(w)$.
Hence we obtain the rate of energy dissipation $\Gamma$ in the thermodynamic limit
\begin{align}
\Gamma&=\lim_{T\to \infty}\frac{{\cal E}_{{\rm tot}}(T)}{T}\\
&=\frac{2\pi n \lambda^2}{ \hbar^2} \int \d3k \;\epsilon _{k} \;\delta(\omega _{k}  -  \v k\cdot \v v).
\end{align}
Thus $\Gamma=0$ unless there exists $\v k$ such that $\omega _{k}  =  \v k\cdot \v v$.
To proceed further, we rewrite the above integral in spherical coordinates and introduce the new variables
$ x= \hbar k/(2Mc)$, and $y= \beta \cos \theta $, where $\beta=v/c$. Then the rate of energy dissipation reads
\be\label{de2}
\Gamma =  \frac{4n \lambda^2M^3c^3}{\pi \hbar^4 \beta} \int_0^{\infty} dx\ x^3 \int _{-\beta} ^{\beta}
dy\ \delta( \sqrt{x^2+1}-  y).
\ee
Clearly it follows that $\beta> 1$ in order to have energy dissipation.
Hence the rate of energy dissipation is
\begin{equation} \label{gamma}
\Gamma = \frac{n \lambda^2M^3c^3}{ \pi \hbar^4 \beta} (\beta^2-1)^2 \;\Theta(\beta-1) ,
\end{equation}
where $\Theta (x)$ is the step function.
Like the Cherenkov radiation, this radiation is emitted inside a cone with an opening angle $ \theta _c = \cos ^{-1} (1/ \beta)$,
and its momentum cuts off at $\hbar k_c=2Mc\sqrt{\beta^2 -1}$ keeping the total dissipated energy finite.
As was mentioned above, this is the microscopic description of Landau's criterion.
We now present the original argument by Landau, which is based on purely kinematical considerations.

Consider any excitation in a fluid, which has atoms of the mass $M$, has an energy $E$
and a momentum $ \v p$ in a reference frame where the fluid is at rest.
Then the energy of the excitation $E '$ in the coordinate system where the fluid is moving with velocity
$\v v$ with respect to the container, is given by a Galilean boost:
$E ' = E + \v p\cdot \v v + M v^2/2$.
Dissipation due to the motion of the excitation can happen if the energy in the boosted frame is negative, i.e.
$ E + \v p \cdot \v v<0$.  In the optimal case where $\v p$ is anti-parallel to $\v v$, this implies $v>E/ p$.
Therefore, for the phonon-like excitation $E =p c$, this condition gives $v>c$, with $c$ the speed of sound in the fluid.
Superfluidity is explained then by the absence of excitations for motions with
velocities smaller than the critical Landau velocity.
While this condition is widely accepted, it seems that it has not yet been derived from an underlying microscopic theory as
was done in this section.

\subsection{Constant Acceleration}
The trajectory of a linear accelerated particle along the $z$-direction is
\be \label{linear}
\v \zeta (t)=\frac{c^2}{a}\sqrt{1+(\frac{at}{c})^2}\;\hat{\v z},
\ee
with a given acceleration $a$ and $c$ the speed of sound.
The speed of the impurity is always bounded by that of sound, $|\dot{\zeta}(t)|<c$.
In the limit $c\to\infty$, the trajectory becomes quadratic in $t$ as $at^2/2+$(constant).
To evaluate the above integral (\ref{int}), we change the integration variable $t$ to a dimensionless
variable $s$ through $t=(c/a)\sinh s$, and we define
\begin{align}\label{Omega}
\Omega_{\v k} &=\frac ca \sqrt{\omega_k ^2- (c \v k \cdot \hat{\v z})^2}=\frac{kc^2}{a}\sqrt{\sin^2 \theta+ k^2\xi^2},\\
\sigma_{\v k} &= \tanh ^{-1} (\frac{c \v k \cdot \hat {\v z}}{\omega_k})
=\tanh ^{-1} (\frac{\cos \theta}{\sqrt{1+ k^2\xi^2}}), \label{sigma}
\end{align}
where $\theta$ is an angle between $\v k$ and the $z$-axis. Then,
\begin{align}
I_{\v k}(\infty) &= \frac{c}{a} \int _{-\infty}^{\infty} ds\;\cosh s\;\exp[i \Omega_{\v k} \sinh (s - \sigma_{\v k})]   \\
&= \frac{2c}{a} [ \pi\;\delta (\Omega_{\v k}) \cosh \sigma_{\v k} + i {\rm K}_1 (\Omega_{\v k}) \sinh \sigma_{\v k}  ],\label{intacc}
\end{align}
where ${\rm K}_1 (x)$ is the modified Bessel function of the second kind.
The first term in eq.~(\ref{intacc}) will not contribute to the occupation number for
the following reasons. The first is that $\Omega_{\v k}=0$ if and only if $\v k=0$.
The second reason follows from two properties of the
Dirac delta distribution; $\delta(f(x))=\sum_i \delta(x-x_i)/|f'(x_i)|$, where
$x_i$ are the roots of $f(x)=0$, and $x\delta(x)=0$.
The final reason is the presence of the factor $\epsilon_k/\omega_k$ in eq.~(\ref{occu}).
Therefore the occupation number is
\be \label{occuacc}
\tilde{n}_{\v k}= \frac{4 c^2 n^2 \lambda ^2}{\hbar^3 a^2 N}
\frac{\epsilon _k}{\omega _k} \v ( {\rm K}_1 (\Omega_{\v k}) \sinh \sigma_{\v k}\v )^2.
\ee
The total dissipated energy is expressed as
\be
{\cal E}_{{\rm tot}}= \frac{4 c^2 n \lambda ^2}{\hbar^2 a^2}
 \int \d3k \epsilon_k\sinh^2 \sigma_{\v k} \;\v ( {\rm K}_1 (\Omega_{\v k})\v )^2.
\ee

Let us examine the above results for the infrared and the ultraviolet region
using the asymptotic behavior of the modified Bessel function.
For this purpose we will take differential forms of the occupation number $d\tilde{n}_{\v k}$ and
the dissipated energy $d{\cal E}_{{\rm tot}}=\hbar \omega_k d\tilde{n}_{\v k}$ in the thermodynamic limit.
From eqs. (\ref{sigma},\ref{occuacc}),
\be \label{dn1}
d \tilde{n}_{\v k}= \frac{2 n \lambda ^2 c^6k^6\cos^2\theta}{(2\pi)^3Ma^4\hbar\omega_k}
\v (\frac{{\rm K}_1 (\Omega_{\v k})}{\Omega _{\v k}}\v )^2 dkdo,
\ee
with $\Omega_{\v k}$ eq.~(\ref{Omega}) and $do$ the element of solid angle.

$\Omega_{\v k}$ becomes very small in the infrared region, and hence the leading term of eq.~(\ref{dn1}) is
\be \label{dn2}
d \tilde{n}_{\v k}\simeq \frac{2 n \lambda ^2 k\cos^2\theta}
{(2\pi)^3Mc^3\hbar\sqrt{1+k^2\xi^2}(\sin^2 \theta+ k^2\xi^2)^2} dkdo.
\ee
Therefore, the leading term has a singularity around $k\simeq 0$ at the angles $\theta=0,\pi$,
i.e. along the direction of trajectory,
which results in the divergent result for the total number of emitted bogolons.
This infrared singularity due to bogolons has the same origin as the famous
infrared catastrophe in quantum electrodynamics \cite{itzykson}.
There, the total number of emitted photons due to a motion of accelerated charge particle diverges as $k\to 0$.
However, the physical measured quantity, the radiated energy, is still finite.

To see the comparison between our case and the infrared catastrophe in quantum electrodynamics,
we first integrate eq.~(\ref{dn2}) over the angles. Then we find that the leading term of $d \tilde{n}_{\v k}$
is $dk/k$, besides a few factors, which diverges for $k \to 0$. However, the total dissipated energy
$d{\cal E}_{{\rm tot}}$ is finite after we multiply $d\tilde{n}_{\v k}$ by $\hbar \omega_k \simeq \hbar kc$.

We next use the asymptotic form of ${\rm K}_1$ for large $\Omega_{\v k}$ to get the leading term:
\be \label{dn3}
d \tilde{n}_{\v k} \simeq \frac{n\lambda^2(2Mc)^3\cos^2\theta}{(2\pi)^2\hbar^5ak^2}\exp(-\frac{\hbar k^2c}{Ma}) dkdo.
\ee
Hence the dissipated energy in the ultraviolet region after integrating over the solid angles is
\be \label{dissacc}
d{\cal E}_{k} \simeq \frac{4n\lambda^2M^2c^3}{ 3\pi \hbar^3a}\exp(-\frac{\hbar k^2c}{Ma}) dk.
\ee
Thus $d{\cal E}_{{\rm tot}}$ asymptotically depends on $k$ through the exponential factor $\exp [-\hbar k^2c/(Ma) ]$ \cite{comment32}.
This high momentum behavior could be used to measure the number $\hbar c/(Ma)$,
from which we can obtain the mass $M$ of a boson.

Lastly we notice that two length scales in this example have different orders of magnitudes.
One is the coherence length $\xi$ and the other is $l_a=c^2/a$.
In most real experimental situations $\xi \ll l_a$, since accelerations available in laboratories
are very restricted. Then we again use the asymptotic form of ${\rm K}_1$
to estimate the total dissipated energy for the weak acceleration limit,
\be
{\cal E}_{{\rm tot}} \simeq \frac{ n \lambda ^2 M^2 c}{10 \hbar ^3} \sqrt{\frac{\hbar a}{\pi M c^3}} .
\ee
Although the speed of impurities never exceed the speed of sound we expect
a small finite amount of dissipation due to the uniformly accelerated impurity.

\subsection{Circular Motion}
The trajectory of relativistic circular motion on the $xy$-plane is discussed in Appendix.
Thus it is enough to consider the following trajectory, given two parameters $R$ and $\Omega>0$.
\be
\zeta_x(t)=R \cos \Omega t,\quad \zeta_y(t)=R \sin \Omega t,\quad\zeta_z(t)=0,
\ee
where the constant term in $\zeta_x(t)$ is dropped since it does not contribute to the dissipated energy.
The integral to evaluate is
\be
I_{\v k}(\infty)= \int_{-\infty}^{\infty} dt\;e^{i\omega_k t-i k_{\bot}R \cos(\Omega t-\phi)},
\ee
where we express $\v k$ in cylindrical coordinates as $\v k =(k_{\bot} \cos \phi, k_{\bot} \sin \phi, k_z)$.
A formula
\be
e^{-ix\cos \theta}=\sum_{\ell=-\infty}^{\infty}(-i)^{\ell}e^{i\ell\theta}{\rm J}_{\ell}(x),
\ee
with ${\rm J}_{\ell}(x)$ the Bessel function, expresses the above integral in a series as
\begin{align}
I_{\v k}(\infty)&=\sum_{\ell}(-i)^{\ell}e^{-i\ell \phi}{\rm J}_{\ell}(k_{\bot}R) \int_{-\infty}^{\infty} dt \;e^{i\omega_k t+i \ell \Omega t}\\
&=2\pi\sum_{\ell}i^{-\ell}e^{i\ell \phi}{\rm J}_{\ell}(k_{\bot}R)\;\delta(\omega_k -\ell\Omega). \label{intcir}
\end{align}
It is clear from the last expression that the above integral vanishes unless
$\omega_k =\ell\Omega$ for some positive integer $\ell$.
When we square (\ref{intcir}) we need to take care of the delta distribution properly in the
same manner as with the linear motion case.
The occupation number for a finite but large time interval $T$ has an asymptotic form:
\be
\tilde{n}_{\v k}  =2\pi T \frac{n^2 \lambda ^2}{\hbar^3  N} \frac{\epsilon _k}{\omega _k}
\sum_{\ell=1}^{\infty}  \delta(\omega_k -\ell\Omega)\v ({\rm J}_{\ell}(k_{\bot}R)\v )^2.
\ee
Hence the total dissipated energy per unit time is
\begin{align}
\Gamma&\equiv\lim _{T\to\infty} \frac{{\cal E}_{{\rm tot}}(T)}{T}\\&= \frac{2 \pi n \lambda ^2}{\hbar^2}\sum_{\ell=1}^{\infty}
 \int \d3k \epsilon_k\;\delta(\omega_k -\ell\Omega) \v ({\rm J}_{\ell}(k_{\bot}R)\v )^2.
\end{align}
To carry out the angle integral we use the formula
\be
\v ({\rm J}_{\ell}(x)\v )^2=\frac{1}{2\pi}\int _{-\pi}^{\pi}d\varphi \;e^{2i\ell\varphi}{\rm J}_0(2x\sin \varphi).
\ee
After several straightforward steps, we get the $\ell$th mode for the rate of energy dissipation $\Gamma$:
\be \label{gamma2}
\Gamma_{\ell}=\Gamma_0 \frac{\ell (k_{\ell}\xi)^2}{1+(k_{\ell}\xi)^2}\sum_{j=0}^{\infty}{\rm J}_{2j+2\ell+1}(2k_{\ell}R),
\ee
where $\Gamma_0=n\lambda^2M\Omega/(2\pi\hbar^2R)$ is a constant with dimension of energy per unit time,
and $k_{\ell}$ is
\be
k_{\ell}\xi=\sqrt{\sqrt{1+2(\frac{\ell\xi\Omega}{c})^2}-1} \quad (\ell =1, 2, \cdots).
\ee
The total dissipated energy is given by summing over all modes $\ell$.

In order to estimate the large $\ell$ behavior of (\ref{gamma2}), we use the method of stationary phase as
discussed before. The stationary phase points are given by
\be
k_{\bot} R \Omega \sin (\Omega \tau_j-\phi)+\omega_k=0.
\ee
Limiting the case $R\Omega/c<1$, i.e. $k_{\bot} R \Omega<\omega_k$, the $\tau_j$ are
\be
\Omega \tau_j-\phi=(\frac 32+2j)\pi\pm i |\cosh^{-1}(\frac{\omega_k}{k_{\bot} R \Omega})| \; (j=0,\pm 1, \cdots).
\ee
We need to choose ${\rm Im} (\tau_j)<0$ to satisfy the condition that the square of the integral will not
diverge for $k\to\infty$. Then, following the procedure discussed before and deforming the contour properly,
we get the asymptotic expression of the integral $I_{\v k}(\infty)$ :
\begin{align}
I_{\v k}(\infty)&\simeq\sum_{j=-\infty}^{\infty}\exp[i\omega_k \tau_j-ik_{\bot}R\cos(\Omega\tau_j-\phi)]
\int dt\;\exp[-\Omega\sqrt{\omega_k^2-(k_{\bot} R\Omega)^2}(t-\tau_j)^2] \\
\nonumber &=(\frac{\pi}{\Omega\sqrt{\omega_k^2-(k_{\bot} R\Omega)^2}})^{1/2}
\exp\{-\frac{1}{\Omega}[\omega_k|\cosh^{-1}(\frac{\omega_k}{k_{\bot} R \Omega})|\\ 
&\hspace{4cm}+\sqrt{\omega_k^2-(k_{\bot} R\Omega)^2}-i\omega_k(\phi+\frac 32\pi)] \}\sum_{j=-\infty}^{\infty}e^{2n\pi i\omega_k/\Omega}\\
\nonumber &=\sum^{\infty}_{\ell=1}(\frac{\pi\Omega}{\sqrt{\omega_k^2-(k_{\bot} R\Omega)^2}})^{1/2}
(\frac{k_{\bot} R\Omega}{\omega_k+\sqrt{\omega_k^2-(k_{\bot} R\Omega)^2}})^{\ell} \\
&\hspace{4cm}\times\exp\{-\frac{1}{\Omega}[\sqrt{\omega_k^2-(k_{\bot} R\Omega)^2}-i\omega_k(\phi+\frac 32\pi)] \}\;\delta(\omega_k-\ell \Omega),
\end{align}
where we used formulae $\sum_{j=-\infty}^{\infty}\exp(ij\theta)=2\pi\sum_{\ell=-\infty}^{\infty}\delta(\theta-2\ell \pi)$
and $\cosh^{-1}x=\ln(x+\sqrt{x^2-1})$.
Thus we have also obtained the same delta distribution dependence as eq.~(\ref{intcir}).
Taking a rough approximation $\sqrt{\omega_k^2-(k_{\bot} R\Omega)^2}\simeq\omega_k$,
\be
I_{\v k}(\infty)\simeq \sum^{\infty}_{\ell=1}
\frac{\sqrt{2\pi\Omega}(k_{\bot} R\Omega)^{\ell}}{(2\omega_k)^{\ell+1/2}}
 e^{-\ell+i\ell(\phi+\frac 32\pi)}\;\delta(\omega_k-\ell \Omega).
\ee
With this result and the relation $\ell \sim \hbar k_{\ell}^2/(2M\Omega)$ for large $\ell$,
we find an exponential dependence of the rate of energy dissipation for large $k$ as $\exp[-\hbar k_{\ell}^2/(M\Omega)]$.
This shows similar behavior as the constant acceleration case where dependence was found as
$\exp[-\hbar k^2 c/(Ma)]$. In most experimental settings the linear acceleration is so weak
that $c/a\gg 1/\Omega$. And hence we expect in the circular motion case, it is easier to estimate
the mass $M$ of bosons by measuring the dissipated energy for high momenta $\hbar k$.

We remark that circular motion in BECs may be easier to realize in laboratories compared to the
constant acceleration case.
One reason is that we have two parameters $R$ and $\Omega$, which when combined
can produce observable dissipation effects. The other reason is that circular motion can be
realized with additional trapping potentials which are necessary in current BEC experiments.
In this context it is also interesting to study the induced density and current profiles (\ref{indden},\ref{indcur})
to examine the validity of our model.

\section{Dynamics of Impurities}
We will consider back reactions on the impurities in this section.
To this end we regard the impurity as a single quantum mechanical particle with
the canonical momentum $\hat{\v p}(t)$ and the position $\hat{\v q}(t)$.
They satisfy the canonical commutation :
\be
[\hat{q}_i, \hat{p}_j]=i\hbar \delta_{ij}\ (i,j=1,2,3).
\ee
We add the Hamiltonian for the impurity $H_{imp}=\hat{\v p}^2/2m$ with $m$ the mass of the impurity.
Then the total Hamiltonian of the system is
\begin{multline}
\hat{H} (t) = \int d^3x\;\hat{\psi}^{\dagger} (\v{ x},t) (-\frac{\hbar ^2 \v{ \nabla} ^2}{2 M})\hat{\psi} (\v{ x},t) \\
+ \frac g2  \int d^3x\;\hat{\psi}^{\dagger} (\v{ x},t) \hat{\psi}^{\dagger}
(\v{ x},t) \hat{\psi} (\v{ x},t) \hat{\psi} (\v{ x},t)\\
+ \lambda  \int d^3x\;\delta(\v x-\hat{\v q}(t)) \hat{\psi}^{\dagger} (\v{ x},t) \hat{\psi} (\v{ x},t)
+\frac{\v( \hat{\v p}(t)\v)^2}{2m}.
\end{multline}
Here we assume that the impurity is moving without an external potential.
The dynamics of the impurity with additional potentials can be found by adding
suitable a potential term for the impurity in the above Hamiltonian.
Following exactly the same procedures as before,
we obtain the Heisenberg equations of motion :
\bal
i\hbar\pd_t\hat{b}_{\v k}(t)&=\hbar \omega_k \hat{b}_{\v k}(t)+f_{\v k}(t),\\
-i\hbar\pd_t\hat{b}^{\dagger}_{\v k}(t)&=\hbar \omega_k \hat{b}^{\dagger}_{\v k}(t)+f^{\dagger}_{\v k}(t),\\
i\hbar \pd_t\hat{\v q}(t)&=i\hbar\hat{\v p}(t)/m,\\
i\hbar \pd_t \hat{\v p}(t)&=\sideset{}{'}\sum \hbar\v k (\hat{b}_{\v k}(t)f^{\dagger}_{\v k}(t)-{\rm h.c.}),
\end{align}
where $f_{\v k}(t)$ now depends on the position operator $\hat{\v q}(t)$, i.e.,
\be \label{f2}
f_{\v k}(t)= n \lambda \sqrt{\frac{\epsilon _{k}}{N \hbar \omega _{k}}}\;\exp(-i\v k\cdot\hat{\v q}(t)).
\ee
Thus we can obtain a closed equation for the position operator $\hat{\v q}(t)$ as
\be \label{impdynamics}
m\ddot{\v q}(t)=-\frac{n\lambda^2}{\hbar^2}\frac{1}{V}\sideset{}{'}\sum
\frac{\hbar \v k\epsilon_k}{\hbar\omega_k}\left[ \exp\v (ig_{\v k}(\v q(t);t)\v )\int_{t_0}^tdt'\;\exp\v(-ig_{\v k}(\v q(t');t')\v )+{\rm c.c.} \right],
\ee
where $g_{\v k}(\v q(t);t)=\omega_k t-\v k \cdot\!\v q(t)$ and the boundary condition is chosen as $t=t_0$.
To derive this equation, we take the expectation value with respect to the homogeneous condensate,
and we drop the hat for the position operators.

Although it is unlikely that we can obtain the analytical solution to the above equation (\ref{impdynamics}),
this equation will be useful to study the dynamics of the impurities numerically.

\section{Alternative Model}
So far we have only considered the local density-density interaction between the impurities and massive bosons.
In this section we provide an alternative interaction model which is Galilean invariant.
We first review Galilean symmetry in the non-relativistic quantum field theory \cite{takahashi86,takahashi88}.

\subsection{Galilean symmetry}
A Galilean transformation with a given velocity $\v v$ is defined by
\bal
t&\to t'=t,\\
\v x&\to \v x'=\v x-\v vt.
\end{align}
Under this Galilean transformation, the partial derivatives $\pd_t$ and $\pd_i$ are transformed as
\bal
\pd_t&\to\pd'_t=\pd_t+v_i\pd_i,\\
\pd'_i&\to\pd_i.
\end{align}
The quantized field $\hat{\psi}(\v x,t)$ is transformed as
\be
\hat{\psi}(\v x,t)\to\hat{\psi'}(\v x',t')=e^{iS_{\v v}(\v x,t)/\hbar}\hat{\psi}(\v x,t),
\ee
where $S_{\v v}(\v x,t)=-M\v v\cdot\!\v x+M\v v^2t/2$.

The Lagrangian density $\cal L$ for the Bogoliubov model,
\be
{\cal L}=\frac{i\hbar}{2}(\hat{\psi}^{\dagger}   \pd_t \hat{\psi}  -\pd_t \hat{\psi}^{\dagger}  \hat{\psi})
-\frac{\hbar^2}{2M}\pd_i \hat{\psi}^{\dagger}   \pd_i \hat{\psi}
-\frac g2 \hat{\psi}^{\dagger}   \hat{\psi}^{\dagger}   \hat{\psi}   \hat{\psi},
\ee
is invariant (scalar) under the Galilean transformation.
Then the corresponding conserved current $N^{\mu}_{\ (i)}$($\mu=0,1,2,3$ and $i=1,2,3$) is given as
\bal
\hat{N}^{0}_{\ (i)}(\v x,t)&=x^i \hat{J}^0(\v x,t)+t \;\hat{T}^{0}_{\ (i)}(\v x,t),\\
\hat{N}^{i}_{\ (j)}(\v x,t)&=x^i \hat{J}^j(\v x,t)+t \;\hat{T}^{i}_{\ (j)}(\v x,t).
\end{align}
Here $\hat{J}^0(\v x,t)=M\hat{\rho}(\v x,t)$ is the mass density, $\hat{J}^i(\v x,t)=M\hat{\jmath}^i(\v x,t)$ is the mass current density, and
the stress energy momentum complex $\hat{T}^{\mu}_{\ (i)}(\v x,t)$ is defined by eqs.~(\ref{tensor}).
For convenience we list them below :
\bal
\hat{J}^0(\v x,t)&=M\hat{\psi}^{\dagger}\hat{\psi},\\
\hat{J}^i(\v x,t)&=\frac{\hbar}{2i}(\hat{\psi}^{\dagger}\pd_i \hat{\psi}-\pd_i \hat{\psi}^{\dagger}\hat{\psi}),\\
\hat{T}^{0}_{\ (0)}(\v x,t)&=\frac{\hbar^2}{2M}\pd_i \hat{\psi}^{\dagger}\pd_i \hat{\psi}
+\frac g2 \hat{\psi}^{\dagger}\hat{\psi}^{\dagger}\hat{\psi}\hat{\psi} ,\\
\hat{T}^{i}_{\ (0)}(\v x,t)&=-\frac{\hbar^2}{2M}(\pd_t \hat{\psi}^{\dagger}\pd_i \hat{\psi}+\pd_i \hat{\psi}^{\dagger}\pd_t \hat{\psi}),\\
\hat{T}^{0}_{\ (i)}(\v x,t)&=\frac{i\hbar}{2}(\hat{\psi}^{\dagger}\pd_i \hat{\psi}  -\pd_i \hat{\psi}^{\dagger}  \hat{\psi}),\\ \nonumber
\hat{T}^{i}_{\ (j)} (\v x,t)&=-\frac{\hbar^2}{2M}(\pd_i \hat{\psi}^{\dagger}  \pd_j \hat{\psi}  +\pd_j \hat{\psi}^{\dagger}  \pd_i \hat{\psi}  )\\ 
&\hspace{0.5cm} +\delta_{ij}[-\frac{i\hbar}{2}(\hat{\psi}^{\dagger}\pd_t \hat{\psi}-\pd_t \hat{\psi}^{\dagger}\hat{\psi}) 
+ \frac{\hbar^2}{2M}\pd_k \hat{\psi}^{\dagger}\pd_k \hat{\psi}  +\frac g2 \hat{\psi}^{\dagger}\hat{\psi}^{\dagger}\hat{\psi}   \hat{\psi} ].
\end{align}
It is straightforward to show that the current satisfies the continuity equation :
\be
\pd_t\hat{N}^{0}_{\ (i)}+\pd_j \hat{N}^{j}_{\ (i)}=0\ (i=1,2,3).
\ee
Therefore, the generator $\hat{G}^i$ defined by
\be
\hat{G}^i =\int d^3x\;\hat{N}^{0}_{\ (i)}(\v x,t),
\ee
is time independent, i.e. $d \hat{G}^i /dt=0$.
The commutator between the field $\hat{\psi}(\v x,t)$ and the generator $\hat{G}^i$ is
\be
[\hat{\psi}(\v x,t), \hat{G}^i]= (Mx^i+t\;i\hbar\pd_i)\hat{\psi}(\v x,t).
\ee

\subsection{Current-current interaction}
We next look for a possible interaction with the impurities based on the Galilean symmetry.
For the purpose of symmetry considerations, we first consider interactions between massive bosons
and a quantized field $\hat{\Psi}(\v x,t)$ for the impurity. The Galilean transformation
is also applied for the impurity field $\hat{\Psi}(\v x,t)$.
The basic idea is to add possible Galilean invariant terms to the Lagrangian density.
The density $\hat{\rho}=\hat{\psi}^{\dagger}\hat{\psi}$ is obviously a scalar under a Galilean transformation.
Hence, the simplest interaction is
\be
{\cal L}_I=-\lambda \hat{\rho} \hat{\rho}_{imp},
\ee
where $\hat{\rho}_{imp}=\hat{\Psi}^{\dagger}\hat{\Psi}$ is the density of the impurity field.
Taking the classical limit, where the density of the impurity is replaced with the Dirac delta distribution $\delta(\v x-\v \zeta(t))$,
we recover the model proposed in Sec.~II.

The next possible interaction is naively expected as the current-current interaction, i.e.
${\cal L}'_I=-\lambda'\hat{\v{\jmath}}\cdot\!\hat{\v{\jmath}}_{imp}$.  However, this interaction is not allowed within
Galilean invariant theory. We note that the current $\hat{\v{\jmath}}$ is transformed under a
Galilean transformation :
\be
\hat{\v{\jmath}}(\v x,t)\to\hat{\v{\jmath}}'(\v x',t')=\hat{\v{\jmath}}(\v x,t)-\v v \hat{\rho}(\v x,t).
\ee
Therefore, the term $\hat{\v{\jmath}}\cdot\!\hat{\v{\jmath}}_{imp}$ is not a scalar under a Galilean transformation.
In order to form a Galilean scalar, we look for a quantity $\hat{\jmath}_4(\v x,t)$ which transforms as
\be
\hat{\jmath}_4(\v x,t)\to\hat{\jmath}_4'(\v x',t')=\hat{\jmath}_4(\v x,t)-\v v\cdot\!\hat{\v{\jmath}}(\v x,t)+\frac 12 \v v^2\hat{\jmath}_4(\v x,t).
\ee
Then, it is easy to show that the following combination
\be
\hat{\v{\jmath}}\cdot\!\hat{\v{\jmath}}_{imp}-\hat{\rho} \hat{\jmath}_{4imp}-\hat{\jmath}_4 \hat{\rho}_{imp},
\ee
is invariant under a Galilean transformation.
We make the following ansatz for $\hat{\jmath}_4(\v x,t)$ :
\be
\hat{\jmath}_4(\v x,t)=-\frac{\hbar^2}{4M^2}(\hat{\psi}^{\dagger}(\v x,t)\v\nabla^2 \hat{\psi}(\v x,t)
+ \v\nabla^2\hat{\psi}^{\dagger}(\v x,t)\hat{\psi}(\v x,t)).
\ee
Another possible form is $\hat{\jmath}_4=\hbar^2\pd_i\hat{\psi}^{\dagger}\pd_i\hat{\psi}/(2M^2)$.
The physical meaning of $\hat{\jmath}_4(\v x,t)$ is clear, if we multiply by the mass $M$.
$M\hat{\jmath}_4(\v x,t)$ is the kinetic energy density.
Thus, we find that the current-current interaction in Galilean invariant theory is
\be
{\cal L}'_I=-\lambda' \left(\hat{\v{\jmath}}\cdot\!\hat{\v{\jmath}}_{imp}-\hat{\rho} \hat{\jmath}_{4imp}-\hat{\jmath}_4 \hat{\rho}_{imp} \right).
\ee
The corresponding Hamiltonian including both ${\cal L}_I$ and ${\cal L}'_I$ is
\be \label{hcurrent}
\hat{H}= \int\!d^3x \;[ \hat{\psi}^{\dagger}(-\frac{\hbar^2\v \nabla^2}{2M}) \hat{\psi}
+\frac g2 \hat{\psi}^{\dagger}\hat{\psi}^{\dagger}\hat{\psi}\hat{\psi} ]\\
+\lambda \int\!d^3x \;\hat{\rho} \hat{\rho}_{imp}
+\lambda'  \int\!d^3x\;\left(\hat{\v{\jmath}}\cdot\!\hat{\v{\jmath}}_{imp}-\hat{\rho} \hat{\jmath}_{4imp}-\hat{\jmath}_4 \hat{\rho}_{imp} \right).
\ee
The classical limit, where a point-like impurity is moving along a given trajectory,
is obtained by the following replacement :
\bal
\hat{\rho}_{imp}(\v x,t) & \to \rho_c(\v x,t)=\delta(\v x-\v \zeta(t)),\\
\hat{\v{\jmath}}_{imp}(\v x,t) &\to \v{\jmath}_c(\v x,t) =\dot{\v \zeta}(t)\delta(\v x-\v \zeta(t)),\\
\hat{\jmath}_{4imp} (\v x,t)&\to \jmath_{4c} (\v x,t)=  \frac 12 \dot{\v \zeta}(t)^2\delta(\v x-\v \zeta(t)).
\end{align}

\subsection{General solution for the current-current interaction}
The general solution within the Bogoliubov approximation is easily obtained for the
Hamiltonian (\ref{hcurrent}) in the same manner as before.
After the Bogoliubov transformation, the Hamiltonian is
\be
\hat{H}\simeq E'_0+\sideset{}{'}\sum \hbar \omega_k \hat{b}^{\dagger}_{\v k}(t)\hat{b}_{\v k}(t)
+\sideset{}{'}\sum(F_{\v k}(t) \hat{b}^{\dagger}_{\v k}(t)+{\rm h.c.}).
\ee
Here $F_{\v k}(t) $ is defined by
\be
F_{\v k}(t)=n\lambda\sqrt{\frac{\epsilon_k}{N\hbar\omega_k}} 
\left[  \tilde{\rho}_{\v k}(t)\vphantom{\frac{\omega_k \v k\cdot\!\tilde{\v \jmath}_{\v k}(t)}{k^2}}\right.\\
\left.+\frac{\lambda'}{\lambda}\,\frac{\omega_k \v k\cdot\!\tilde{\v \jmath}_{\v k}(t)}{k^2}
-\frac{\lambda'}{\lambda}\,(\tilde{\jmath}_{4\v k}(t)+\frac{\epsilon_k}{2M}\tilde{\rho}_{\v k}(t)) \right].
\ee
Fourier transforms of $\rho_c(\v x,t)$, $\v{\jmath}_c(\v x,t) $, and $\jmath_{4c} (\v x,t)$ are
denoted by $ \tilde{\rho}_{\v k}(t)$, $\tilde{\v \jmath}_{\v k}(t)$, and $\tilde{\jmath}_{4\v k}(t)$, respectively, i.e.,
\bal
\tilde{\rho}_{\v k}(t)&=\int d^3x\;\rho_c(\v x,t)e^{-i\v k\cdot\v x},\\
\tilde{\v \jmath}_{\v k}(t)&=\int d^3x\;\v{\jmath}_c(\v x,t)e^{-i\v k\cdot\v x},\\
\tilde{\jmath}_{4\v k}(t)&=\int d^3x\;\jmath_{4c} (\v x,t)e^{-i\v k\cdot\v x}.
\end{align}
By solving the Heisenberg equation of motion, we obtain
\begin{align}
\hat{b}^{\dagger}_{\v k}(t)&=\hat{b}^{\in\dagger}_{\v k}(t)+\Phi ^*_{\v k}(t),\\
\hat{b}_{\v k}(t)&=\hat{b}^{\in}_{\v k}(t)+\Phi _{\v k}(t) .
\end{align}
Here $\Phi _{\v k}(t)$ is \be \Phi _{\v k}(t)  =  -
\frac{i}{\hbar} e^{-i \omega _{k} t} \int ^t _{-\infty} dt'\;F_{\v
k}(t')e^{i \omega _{k} t'}. \ee The same analyzes as were done before, will hold for the above general solution by replacing
$\phi _{\v k}(t)$ with $\Phi _{\v k}(t) $.

\section{Conclusion and Discussion}
We have shown that the motion of classical impurities in the homogeneous BEC creates the time dependent coherent state. 
This results in macroscopic phenomena such as energy dissipation, induced density fluctuations, and so on. 
The essence of our results is that the homogeneous condensate is very robust against
the motion of the impurities in the large $N$ limit.  
However, in the current experiments where the typical number of atoms are on the order of $10^{5\sim 7}$, 
the motion of the impurities could destroy the condensates by increasing the number of impurities. 
In order to study these effects, we need numerical analysis with additional trapping potentials. 
An extension to the finite temperature is also desirable for this purpose.

We have also shown that our simple model revealed various properties of the homogeneous condensate 
which have not been discussed before. These properties support the possibility of BECs being the space-time medium for impurities. 
Minkowski space-time structure was shown to be a certain part of the homogeneous BEC in the low energy scale, such as 
the wave equations (\ref{waveeq2}), an effective Minkowski metric (\ref{eta1}) and its inverse (\ref{eta2}).   
In this context special relativity could be considered as an effective theory rather than a fundamental one, 
which emerges from the underlying microscopic degrees of freedom.  
Then it is quite natural to expect that the Lorentz symmetry is not the exact symmetry at all energy scales. 
We again emphasize the non-triviality of our result regarding the energy momentum stress complex. 
The deviation of this quantity from the homogeneous condensate 
is shown to be symmetric with an effective Minkowski metric (\ref{eta1}). 


Several examples worked out in Sec.~V illustrate the significance of our model. 
The effective attractive Yukawa type force between stationary impurities in our model 
is understood as the exchange of bogolons whose dispersion relation is linear.  
This effect itself seems difficult to observe directly with the current limitations in experimental techniques. 
Because this force decays exponentially for distance longer than the coherence length. 
The microscopic derivation of Landau's criterion shows that this criterion is not 
so strong one as was originally proposed. It is found that there occur finite amount of energy dissipation 
even slightly below the critical velocity given by eq. (\ref{criticalvel}). However, this dissipation decays 
exponentially for these subluminal motion of impurities. 
We found that energy dissipation for the uniformly accelerated motion of impurity is too small 
to observe by direct measurements. 
In stead we suggest to look for a circular motion case in current experiments. 
Further investigations for this case need to be done with numerical analysis. 

There are several future directions in studying this problem. 
First, we need to treat the impurity at the quantum level rather than as a point-like classical object.  
As we have already seen, the dynamics of a single quantum point-like impurity is difficult to solve analytically.  
Another possible extension of our model is to consider a quantized field for the impurities as was outlined in the previous section. 
Then, we solve the field equations for massive bosons and the impurities self-consistently.  
Another issue which was not studied in our model is inhomogeneous condensates. 
Particularly, impurity effects in the presence of quantized vortices is of interest. 
We would like to continue to study these problems in near future. 

We finally remark that our result does not agree with the conventional approach to quantum field theory in 
curved space-time \cite{unruh}. According to this work, a uniformly accelerated particle detector in the empty Minkowski space-time is 
shown to detect a thermal spectrum, which leads to an interpretation of the thermal effect.   
If we follow the idea proposed in refs. \cite{mazur0,mazur1,mazur2,mazur3,mazur4,mazur5,mazur6,laughlin,chapline} 
and if there exists the Unruh effect, then the spectrum of bogolon for the uniformly accelerated impurity case should be the thermal one. 
As we have shown in our model, the bogolon spectrum for the uniformly accelerated impurity 
in the homogeneous condensate is not the thermal spectrum (eq.~(\ref{occuacc}) or (\ref{dn1})).  
We note that similar, but phenomenological, considerations were presented 
in Chapter 6 of ref.~\cite{volovik1} \cite{volovik}. The author of this book claims that the excitations 
caused by the uniformly accelerated relativistic motion of a point-like 
classical impurity in a superfluid $^3$He are emitted with the thermal spectrum.  
His argument is solely based on the simple Boltzmann factor evaluated in the WKB approximtion. 
The presence of this Bolzmann factor, not a Bose-Einstein factor, for the 
transition rate leads that author to the conclusion that an analog of the 
Unruh effect is present in the B-phase of $^3$He. However, we would disagree 
with the interpretation that the homogeneous condensate would serve as a 
thermal bath for accelerated impurities. 
The reason is clear as seen below \cite{lee}. 
The thermal nature of reservoir is a consequence of a statistical ensemble average. 
From this averaging procedure, the expectation value of occupation number for bosons results in the Plankian distribution, 
but not the other way around.  In other words, an apparent mathematical expression {\it does not} mean 
that the system is in a thermal equilibrium at certain temperature.  
This point was also discussed by Lee within the context of the Unruh effect \cite{lee}. 
In our model the homogeneous condensate is seen as a coherent state by the bogolon, i.e. eq. (\ref{result}).  
Hence, the occupation number (\ref{occu}) with respect to the homogeneous condensate 
has nothing to do with any statistical ensemble average.  This observation leads us to 
the main conclusion that the homogeneous condensate does not serve as a thermal bath for 
accelerated impurities. If the homogeneous condensate is the 
physically correct model for Minkowski space-time, 
it then logically follows that the apparent thermal response of the linearly 
accelerated detector models may be the result of improper regularization.

\begin{acknowledgments}
The author would like to express his acknowledgment to Professor Pawel O. Mazur for
suggesting this problem, many valuable discussions, 
explaining the physical basis of refs. \cite{mazur0, mazur1, mazur2, mazur3, mazur4}, 
and providing the additional information about these references.
He would also like to thank Prof. J.~Knight and Dr. Z.~P.~Karkuszewski for useful comments on the draft. 
This work was supported in part by the National Science Foundation, Grant No. PHY-0140377,
to the University of South Carolina. 
\end{acknowledgments}

\appendix
\section{Relativistic Circular Motion}
We look for a relativistic circular motion in $(2+1)$ space-time. 
We assume that the four acceleration $w^{\mu}(s)=d^2 x^{\mu}(s)/ds^2$ ($\mu=0,1,2$) is of the form: 
\be
w^0 (s)=0,\ \ddot{Z}(s)=w^1(s)+iw^2(s)=\frac{a}{c^2}e^{i \varphi(s)}, 
\ee
where $s$ is the proper time, $a$ is an acceleration constant, and $\varphi(s)$ is 
a real function of the proper time.  Thus $w^{\mu}w_{\mu}=-a^2/c^4$ and $w^{\mu}$ is a space-like. 
The four velocity $u^{\mu}(s)=dx^{\mu}(s)/ds$ satisfies $u^{\mu}u_{\mu}=1$. 
We solve this differential equation with the following initial condition, 
\be
Z(s)|_{s=0}=R_0,\quad \dot{Z}(s)|_{s=0}=i R_0\Omega_0 \left. \frac{dx^0(s)}{ds}\right |_{s=0}, 
\ee
with given radius $R_0$ and angular velocity $\Omega_0$ corresponding to the non-relativistic limit.  
The relation between the acceleration $a$ and these parameters 
$R_0$, $\Omega_0$ is $a=R_0\Omega_0^2$. 
The solution is 
\begin{align}
x^0(s)&=c t(s)=\gamma s,\\ Z(s)&=x(s)+iy(s)=R_0\gamma^2\exp(i\frac{\beta s}{R_0 \gamma})-R_0\gamma^2\beta^2, 
\end{align}
where $\beta=R_0 \Omega_0/c$ and $\gamma= 1/\sqrt{1-\beta^2}$ the relativistic factor. 
From this solution we find the trajectory of relativistic circular motion in terms of time $t$ as 
\be
x(t)=R_0\gamma^2[\cos(\frac{\Omega_0 t}{\gamma^2})-\beta^2],
\ y(t)=R_0\gamma^2\sin(\frac{\Omega_0 t}{\gamma^2}).
\ee
In the limit $c\to\infty$, this solution becomes 
\be
x(t)=R_0 \cos(\Omega_0 t),\ y(t)=R_0\sin(\Omega_0 t).
\ee
We remark the center of the circular motion is not fixed at the origin before the limit $c\to\infty$.



\begin{thebibliography}{99}
\bibitem{einstein} A.~Einstein, Sitzungsberichte, Preussische Akademie der Wissenschaften, 261 (1924); {\it ibid.}, 3 (1925).
\bibitem{bec1} M.~H.~Anderson, J.~R.~Ensher, M.~R.~Matthew, C.~E.~Wieman, and E.~A.~Cornell, Science {\bf 269}, 198 (1995).
\bibitem{bec2} C.~C.~Bradley, C.~A.~Sackett, J.~J.~Tollett, and R.~G.~Hulet, Phys. Rev. Lett. {\bf 75}, 1687 (1995).
\bibitem{bec3} K.~B.~Davis, M.~-O.~Mewes, M.~R.~Andrews, N.~J.~van Druten, D.~S.~Durfee, D.~M.~Kurn, and W.~Ketterle, Phys. Rev. Lett. {\bf 75}, 3969 (1995).
\bibitem{bec} For a review, e.g., F.~Dalfovo, S.~Giorgini, L.~P.~Pitaevskii, and S.~Stringari, Rev. Mod. Phys. {\bf 71}, 463 (1999).
\bibitem{analog} For instance see \cite{volovik2,visser} and references therein.
\bibitem{mazur0} P.~O.~Mazur, Acta Phys. Polon. {\bf 27}, 1849 (1996); hep-th/9603014.
\bibitem{mazur1} P.~O.~Mazur, ``Gravitation as a Many Body Problem", in {\it Beyond the Standard Model V}, 
ed. by G.~Eigen {\it et al}, Woodbury, AIP (1997); hep-th/9708133.
\bibitem{mazur2} P.~O.~Mazur, ``On Gravitation and Quanta", 
(An abstract submitted to the 8th Marcel Grossmann Meeting, Jerusalem 1997); hep-th/9712208.  
(This reference discusses the constituent model of gravitation, condensates, 
and the nonvanishing positive vacuum energy as the finite size effect following \cite{mazur0}.)
\bibitem{mazur3} A.~Z.~G\'{o}rski and P.~O.~Mazur, ``The Quantum Black Hole Specific Heat is Positive", unpublished; hep-th/9704179.
\bibitem{mazur4} P.~O.~Mazur and E.~Mottola, ``Gravitational Condensate Stars: An Alternative to Black Holes", 
unpublished but submitted to Phys. Rev. Lett. in early September 2001; gr-qc/0109035. 
(The first version of this paper was originally written in January 1998 soon after refs. \cite{mazur0,mazur1,mazur2,mazur3} have appeared.) 
\bibitem{mazur5} P.~O.~Mazur and E.~Mottola, Proc. Nat. Acad. Sci.  {\bf 111}, 9545 (2004).
\bibitem{mazur6} G.~Chapline and P.~O.~Mazur, ``Superfluid Picture for Rotating Space-Times", preprint; gr-qc/0407033. 
\bibitem{laughlin} R.~B.~Laughlin, Int. J. Mod. Phys. A {\bf 18}, 831 (2003).
\bibitem{chapline} G.~Chapline, E.~Hohlfeld, R.~B.~Laughlin and D.~I.~Santiago, Int. J. Mod. Phys. A {\bf 18}, 3587 (2003).
\bibitem{dirac1} P.~A.~M.~Dirac, Nature {\bf 168}, 906 (1951); {\it ibid}. {\bf 169}, 146 and 702 (1952).
\bibitem{dirac2} P.~A.~M.~Dirac, Scientific Monthly {\bf 78}, 142 (1954).
\bibitem{michelson} H.~M\"uller, S.~Herrmann, C.~Braxmaier, S.~Schiller, and A.~Peters, Phys. Rev. Lett. {\bf 91}, 020401 (2003).
\bibitem{volovik1} G.~E.~Volovik, {\it Exotic properties of superfluid $^3$He}, World Scientific Publisher (1992). 
\bibitem{volovik2} G.~E.~Volovik, {\it The universe in a helium droplet}, Oxford University Press (2003).
\bibitem{visser} C.~Barcel\'o, S.~Liberati, and M.~Visser, Class. Quantum. Grav. {\bf 18}, 1137 (2001).
\bibitem{meystre} W.~Zhang, H.~Pu, C.~P.~Search, P.~Meystre, and E.~M.~Wright, Phys. Rev. A {\bf 67}, 021601(R) (2003).
\bibitem{kovrizhin} D.~L.~Kovrizhin and L.~A.~Maksimov, Phys. Lett. A {\bf 282}, 421 (2001).
\bibitem{ms} P.~O.~Mazur and J.~Suzuki, unpublished (2003).
\bibitem{astracharchik} G.~E.~Astrakharchik and L.~P.~Pitaevskii, Phys. Rev. A {\bf 70}, 013608 (2004).
\bibitem{mazets} I.~E.~Mazets and G.~Kurizki, in {\it Decoherence, Entanglement and Information
Protection in Complex Quantum Systems}, ed. by V.~M.~Akulin {\it et al}, Springer (2005); cond-mat/0401172.
\bibitem{timmermans} E.~Timmermans and R.~C\^ot\'e, Phys. Rev. Lett. {\bf 80}, 3419 (1998).
\bibitem{jun} J.~Suzuki, unpublished; cond-mat/0407714.
\bibitem{bogoliubov} N.~N.~Bogoliubov, J. Phys. (Moscow) {\bf 11}, 23 (1947).
\bibitem{bogoliubov2} N.~N.~Bogoliubov, {\it Lectures on Quantum Statistics vol. 1}, Gordon and Breach (1968).
\bibitem{bru} See also recent review on the Bogoliubov model, V.~Zagrebnov and J.~-B.~Bru, Phys. Rep. {\bf 350}, 291 (2000).
\bibitem{unruh} W.~G.~Unruh, Phys. Rev. D {\bf 14}, 870 (1976).
\bibitem{unruh2} S.~A.~Fulling and W.~G.~Unruh, Phys. Rev. D {\bf 70}, 048701 (2004).
\bibitem{belinskii} V.~A.~Belinskii, B.~M.~Karnakov, V.~D.~Mur, and N.~B.~Narozhnyi, JETP Lett. {\bf 65}, 902 (1997).
\bibitem{belinskii2} N.~B.~Narozhny, A.~M.~Fedotov, B.~M.~Karnakov, V.~D.~Mur, and V.~A.~Belinskii, Phys. Rev. D {\bf 65}, 025004 (2001).
\bibitem{belinskii3} N.~B.~Narozhny, A.~M.~Fedotov, B.~M.~Karnakov, V.~D.~Mur, and V.~A.~Belinskii, Phys. Rev. D {\bf 70}, 048702 (2004).
\bibitem{comment31} The original Bogoliubov solution was obtained for the general two-body interaction case in 1947 \cite{bogoliubov}.
\bibitem{comment2} There are several proposals for the number conserving treatment for BEC 
\cite{bogoliubov2, girardeau, sasaki, gardiner, castin}. 
We follow the one used in refs.~\cite{bogoliubov2, girardeau}.
\bibitem{girardeau} M.~D.~Girardeau and R.~Arnowitt, Phys. Rev. {\bf 113}, 755 (1959); 
M.~D.~Girardeau, Phys. Rev. A {\bf 58}, 775 (1998). 
\bibitem{sasaki} S.~Sasaki and K.~Matsuda, Prog. Theor. Phys. {\bf 56}, 375 (1975).
\bibitem{gardiner} C.~W.~Gardiner, Phys. Rev. A {\bf 56}, 1414 (1997).
\bibitem{castin} Y.~Castin and R.~Dum, Phys. Rev. A {\bf 57}, 3008 (1998).
\bibitem{umezawa1} H.~Umezawa, H.~Matsumoto, and M.~Tachiki, {\it Thermo Field Dynamics and Condensed States}, North-Holland (1982).
\bibitem{umezawa2} H.~Umezawa, {\it Advanced Field Theory--Micro, Macro and Thermal Physics}, AIP (1993).
\bibitem{fetter} See for instance; A.~L.~Fetter and J.~D.~Walecka, {\it The Quantum Theory of Many Particle Systems}, Dover (2003).
\bibitem{gross} E. P. Gross, in {\it Mathematical Methods in Solid State and Superfluid Theory},
R. C. Clark and G. H. Derrick ed. Oliver \& Boyd, Edinburgh (1969).
\bibitem{preparata} G.~Preparata, {\it QED Coherence in Matter}, World Scientific Publisher (1995).
\bibitem{landau} L.~D.~Landau, J. Phys. (Moscow) {\bf 5}, 71 (1941).
\bibitem{itzykson} C.~Itzykson and J-B.~Zurber, {\it Quantum Field Theory}, McGraw-Hill (1980).
\bibitem{stoof} M.~J.~Bijlsma, B.~A.~Heringa, and H.~T.~C.~Stoof, Phys. Rev. A {\bf 61}, 053601 (2000).
\bibitem{pethick} L.~Viverit, C.~J.~Pethick, and H.~Smith, Phys. Rev. A {\bf 61}, 053605 (2000).
\bibitem{comment32} This factor can also be obtained by evaluating the integral using the stationary phase method
explained in Sec.~V.  We would like to thank Prof. P. O. Mazur for pointing this out.
\bibitem{takahashi3} Y.~Takahashi, {\it Introduction to Field Analysis}, Koudan-sha (1982) (in Japanese).
\bibitem{takahashi86} Y.~Takahashi, in {\it Rationale of Beings}, ed. by K.~Ishikawa et al., World Scientifc Publisher (1986).
\bibitem{takahashi88} Y.~Takahashi, Fortschr. Phys. {\bf 36}, 63 and 83 (1988).
\bibitem{volovik} We thank Dr. G.~E.~Volovik for providing information on his book \cite{volovik1}.
\bibitem{lee} T.~D.~Lee, Nucl. Phys. B {\bf 264}, 437 (1986).
\end{thebibliography}
\end{document}